%% file: main.tex
\documentclass[%
 reprint,
 superscriptaddress,
nofootinbib,
 amsmath,amssymb,
 aps,
longbibliography,
floatfix,
]{revtex4-2}

\usepackage{graphicx}%
\usepackage{dcolumn}%
\usepackage{bm}%

\usepackage[utf8]{inputenc} %
\usepackage[T1]{fontenc}    %
\usepackage[
bookmarks=true,
colorlinks=true,
citecolor=red,
linkcolor=blue,
pagebackref=true]{hyperref}
\usepackage[normalem]{ulem}
\usepackage{url}            %
\usepackage{booktabs}       %
\usepackage{amsfonts}       %
\usepackage{nicefrac}       %
\usepackage{mathtools}
\usepackage{xcolor}         %
\usepackage[braket, qm]{qcircuit}
\definecolor{lightred}{rgb}{0.749,0.196,0.447}
\definecolor{customyellow}{rgb}{1.0, 1.0, 0.6}
\definecolor{lightpink}{rgb}{1.0, 0.9, 0.9}
\definecolor{FireBrick}{rgb}{0.698,0.133,0.133}
\definecolor{RoyalBlue}{rgb}{0.255,0.41,0.884}
\definecolor{DarkerFireBrick}{rgb}{0.4886, 0.0931, 0.0931}
\definecolor{DarkerRoyalBlue}{rgb}{0.1785, 0.287, 0.6188}

\newcommand{\ours}{L2O-$g^{\dagger}$\xspace}

\usepackage{fontawesome5}
\usepackage{anyfontsize}

\usepackage{amsmath, amssymb, amsthm}
\usepackage{physics}
\usepackage{nicefrac}

\usepackage{graphicx}
\usepackage{booktabs}
\usepackage{multirow}
\usepackage{array}
\usepackage{diagbox}
\usepackage[justification=raggedright,singlelinecheck=true]{caption}
\usepackage{subcaption}
\usepackage{wrapfig}

\usepackage[capitalize]{cleveref}
\usepackage{bookmark}

\usepackage{enumitem}
\usepackage{mdframed}
\usepackage{placeins}
\usepackage{rotating}
\usepackage{soul}
\usepackage{ulem}

\usepackage{algorithm}
\usepackage{algpseudocode}
\usepackage{xspace}
\usepackage{qcircuit}
\usepackage{tikz}
\usetikzlibrary{fit}
\usetikzlibrary{arrows.meta, shapes, positioning, shadows.blur, trees, backgrounds}
\usepackage{lipsum, zhlipsum}
\usepackage{xcolor}
\usepackage{transparent}
\usepackage{float}
\usepackage{color, colortbl}
\usepackage{tcolorbox}
\usepackage{braket}
\usepackage{adjustbox}
\usepackage{tabularx}

\theoremstyle{remark}

\theoremstyle{plain}

\theoremstyle{plain}

\begin{document}

\preprint{APS/123-QED}

\title{\ours: Learning to Optimize Parameterized Quantum Circuits\\ with Fubini–Study Metric Tensor}

\author{Yu-Chao Huang}
\email{r11222015@ntu.edu.tw}
\affiliation{
  Department of Physics and Center for Theoretical Physics, National Taiwan University, Taipei 106319, Taiwan
}

\author{Hsi-Sheng Goan}
\email{goan@phys.ntu.edu.tw}
\affiliation{
  Department of Physics and Center for Theoretical Physics, National Taiwan University, Taipei 106319, Taiwan
}
\affiliation{
  Center for Quantum Science and Engineering, National Taiwan University, Taipei 106319, Taiwan
}
\affiliation{
  Physics Division, National Center for Theoretical Sciences, Taipei 106319, Taiwan
}

\date{\today}%

\input{Tex/0_abstract}

\maketitle

\input{Tex/1_Introduction}

\input{Tex/3_Theory}

\input{Tex/2_Methodology}

\input{Tex/5_Experiments}

\input{Tex/4_Related}

\input{Tex/6_Conclusion}

\begin{acknowledgments}
H.-S.G. acknowledges support from the National Science and Technology Council, Taiwan under Grants No.~NSTC 113-2119-M-002 -021, No.~NSTC112-2119-M-002-014, No.~NSTC 111-2119-M-002-007, and No.~NSTC 111-2627-M-002-001, from the US Air Force Office of Scientific Research under Award Number FA2386-20-1-4052, and from the National Taiwan University under Grants No.~NTU-CC-112L893404 and No.~NTU-CC-113L891604. H.-S.G. is also grateful for the support from the ``Center for Advanced Computing and Imaging in Biomedicine (NTU-113L900702)'' through The Featured Areas Research Center Program within the framework of the Higher Education Sprout Project by the Ministry of Education (MOE), Taiwan, and the support from the Physics Division, National Center for Theoretical Sciences, Taiwan.
\end{acknowledgments}

\clearpage
\bibliographystyle{plainnat}
\bibliography{Bib/references, Bib/custom}

\clearpage
\appendix
\input{Tex/Appendix}

\end{document}

%% file: Tex/0_abstract.tex
\begin{abstract}
Before the advent of fault-tolerant quantum computers, variational quantum algorithms (VQAs) play a crucial role in noisy intermediate-scale quantum (NISQ) machines.
Conventionally, the optimization of VQAs predominantly relies on manually designed optimizers.
However, learning to optimize (L2O) demonstrates impressive performance by training small neural networks to replace handcrafted optimizers. 
In our work, we propose \ours, a \textit{quantum-aware} learned optimizer that leverages the Fubini-Study metric tensor ($g^{\dagger}$) and long short-term memory networks.
We theoretically derive the update equation inspired by the lookahead optimizer and incorporate the quantum geometry of the optimization landscape in the learned optimizer to balance fast convergence and generalization. 
Empirically, we conduct comprehensive experiments across a range of VQA problems. Our results demonstrate that \ours not only outperforms the current SOTA hand-designed optimizer without any hyperparameter tuning but also shows strong out-of-distribution generalization compared to previous L2O optimizers. We achieve this by training \ours on just a single generic PQC instance.
Our novel \textit{quantum-aware} learned optimizer, \ours, presents an advancement in addressing the challenges of VQAs, making it a valuable tool in the NISQ era. 
The implementation and data are available at: \url{https://github.com/Physics-Morris/L2O-g}.
\end{abstract}

%% file: Tex/1_Introduction.tex
\section{Introduction}

Quantum computers provide a different paradigm of computing based on quantum mechanical principles such as superposition and entanglement.
It has the potential to change the future of computing~\cite{Shor_1997, daley2022practical}. 
However, a reliable quantum computer for tasks like factoring large numbers requires $10^6$ qubits and a probability of error per quantum gate of the order of $10^{-6}$ \cite{Preskill_1998}. 
Currently, we have around $10^2$ qubits and an error rate between $10^{-2}$ and $10^{-4}$\cite{google2023suppressing}.
Before the advent of the first fault-tolerant quantum computer, noisy intermediate-scale quantum (NISQ) computing~\cite{preskill2018quantum} with hundreds of qubits could potentially surpass classical computing. 
In particular, variational quantum algorithms (VQAs) emerge as crucial algorithms for NISQ devices \cite{Cerezo_2021}. 
In essence, VQAs consist of a parameterized quantum circuit (PQC) and use a classical computer to optimize the cost function based on the output of the quantum computer (see~\Cref{fig:opt_diagram}).
Applications include many-body quantum physics~\cite{PRXQuantum.2.030346}, chemistry~\cite{McArdle_2020}, combinatorial optimization~\cite{farhi2014quantum, PhysRevResearch.2.023302}, and machine learning~\cite{bharti2021noisy, Schuld_2020}.

\begin{figure}[htp!]
    \vspace{-1em}
    \includegraphics[scale=0.40]{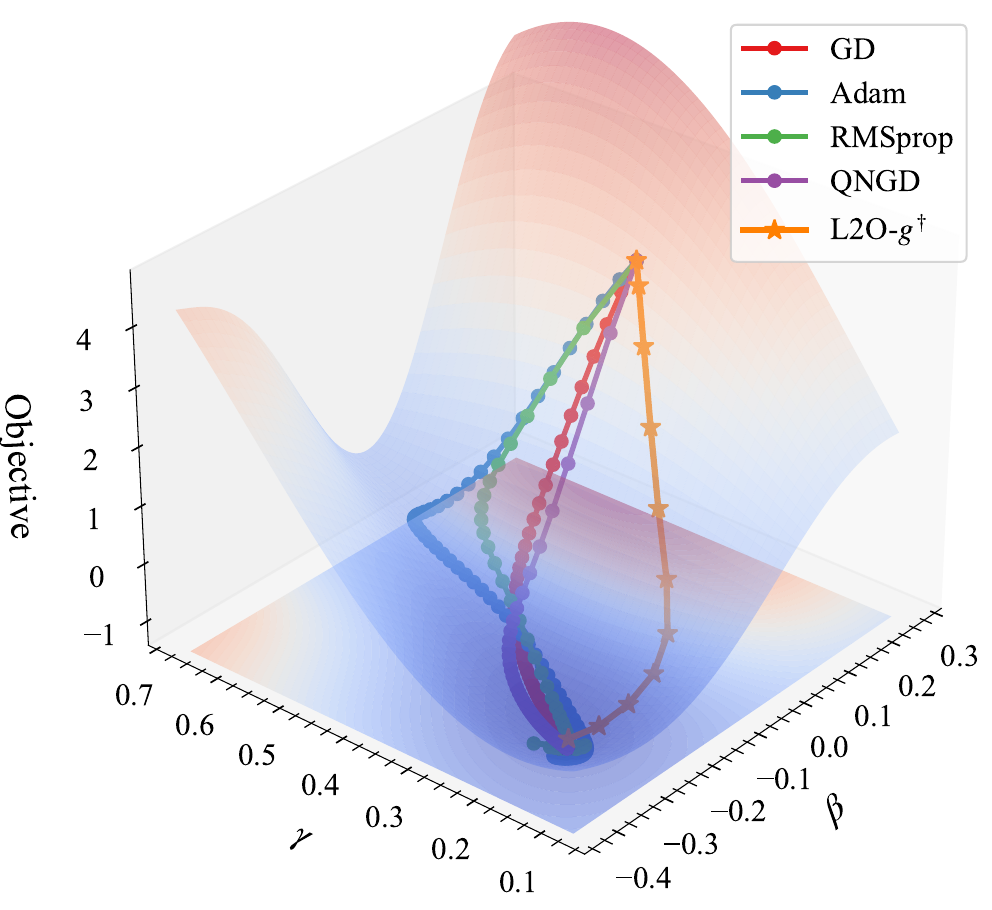}{}
    \vspace{-1em}
    \caption{\small
    \textbf{A Motivating Toy Example of \ours Optimizing MaxCut Problem with QAOA ($V=5, p=0.9, p_{\text{layer}}=1$).} 
    \ours automatically adjusts step size approaching the minimum and converges to the minimum with only a few steps.}
    \label{fig:loss_traj}
\end{figure}

However, it remains uncertain whether VQAs can surpass classical computers in the NISQ era as they scale up.
On the classical optimization side, \citet{PhysRevLett.127.120502} show that even if quantum systems are classically tractable (e.g., logarithmic qubits or free fermions), classical optimization remains $\mathsf{NP}$-hard. 
Furthermore, increasing the expressiveness of the variational ansatz often leads to trainability issues like barren plateaus and narrow gorges \cite{Cerezo_2021_cost}.
\citet{mcclean2018barren} show that the exponential vanish of the gradient with the number of qubits makes it untrainable.
To circumvent the problem, there are three active areas: parameters initialization technique \cite{Grant_2019, zhang2022escaping}, circuit design \cite{liu2022mitigating, grimsley2023adaptive, Cerezo_2021}, and optimization technique \cite{Stokes_2020, Gacon_2021, Magann_2022, Koczor_2022, schuld2019evaluating}.
\ours falls into the last category, \ours improving the optimization with an auto-adjustable training strategy consisting of a quantum-aware learned optimizer (see toy example in \Cref{fig:loss_traj}).

When optimizing a PQC in VQAs, we usually use a hand-designed analytical algorithm. 
Additionally, most optimizer requires a careful choice of optimizer and global search for hyperparameters given different optimizing problems. 
To address this issue, learning to optimize (L2O) uses a small neural network to replace the handcrafted optimizer to learn the optimal optimization strategy tailored for a similar distribution of tasks, and dynamically adjust during the optimization process.
This approach leverages the intuition of a neural network as a universal function approximator~\cite{hornik1989multilayer}.

Leaned optimizer was introduced by \citet{andrychowicz2016learning} which uses a coordinate-wise LSTM to replace hand-crafted optimizer.
Their results show faster convergence compared to hand-crafted optimizers on similar classes of optimization problems in their training sets.
Inherit the same architecture from \citet{andrychowicz2016learning}, \citet{verdon2019learning} and \citet{wilson2019optimizing} demonstrate the simple LSTM learned optimizer's ability to tackle the problem of optimizing VQAs under limited scenarios away from the training distribution.
However, \textit{No Free Lunch Theorem}~\cite{poland2020free} suggest there is no single optimizer that trumps others on every objective.
Therefore, we set out to address the generalizability problem of previous purely classical learned optimizers by replacing the hand-designed optimizer with a \textit{quantum-aware} learned optimizer tailored to tackle optimization in VQAs.

In this work, we propose a \textit{quantum-aware} learned optimizer, \ours, which dynamically adjusts the balance between parameter space optimization and distribution space optimization during the optimization process. We theoretically derive the update equation to incorporate the quantum geometry of the parameter space in the learned optimizer, using the Fubini-Study metric tensor along with learned update directions and update steps to effectively balance fast convergence and generalization across different VQAs. Specifically, a learnable vector balances the parameter space and distribution space optimization coordinate-wise. We show that \ours generalizes to all kinds of optimization problems in VQAs by training on a single generic PQC instance. Without any hyperparameter tuning, it matches or surpasses a hyperparameter-tuned hand-crafted optimizer in optimizing a wide variety of VQAs. Limitations of our work are discussed in \Cref{sec:limit}.

\subsection{Contributions}
Our contributions are summarized as follows:
\begin{itemize}
    \item We propose \ours with theoretical motivation and empirically demonstrate its robustness against a diverse array of optimization problems in VQAs.
    \item We show that \ours can generalize well to a wide range of VQAs problems by training on a single generalized PQC instance.
    \item We open source our implementation, experiments data, and results for minimum effort integration of any future VQAs problems with \ours.
\end{itemize}

\subsection{Notation}
We denote the set $[n] \coloneqq \{1, \dots, n\}$.
We denote vectors in lowercase bold letters and matrices in uppercase bold letters.
The Pauli matrices are denoted by $\hat{\sigma}_x$, $\hat{\sigma}_y$, and $\hat{\sigma}_z$. The operator norm of an operator $\hat{X}$ is denoted by $\|\hat{X}\|$.
The \(\ell_2\)-norm of a vector $\mathbf{x}$ is denoted by $\|\mathbf{x}\|_2$.
The inner product of vectors $\mathbf{x}$ and $\mathbf{y}$ is denoted by $\langle \mathbf{x}, \mathbf{y} \rangle$.
The element-wise product of vectors $\mathbf{x}$ and $\mathbf{y}$ is denoted by $\mathbf{x} \circ \mathbf{y}$ and element-wise application of a function \(f\) to a vector $\mathbf{x}$ is denoted by $f[\mathbf{x}]$.

\input{figures/opt_diagram}

%% file: figures/opt_diagram.tex
\begin{figure}[htb]
\begin{tikzpicture}[node distance=0.8cm, auto, scale=0.75, transform shape]

\definecolor{color1}{HTML}{CBC7FC}
\definecolor{color2}{HTML}{FFED86}
\definecolor{color3}{HTML}{A2DCE7}
\definecolor{color4}{HTML}{F8CCDC}
\definecolor{color5}{HTML}{E7F2F8}

  \node (input) [rectangle, draw, fill=color1, text width=4cm, text centered, rounded corners, minimum height=3em] {Input $\{[\rho_D], U(\bm{\theta}), |\Psi_0\rangle, \mathcal{C}(\bm{\theta})\}$};

  \foreach \i in {1,...,3}
    {
      \pgfmathsetmacro{\shift}{(\i - 1) * 0.1}
      \node (quantum\i) [rectangle, draw, fill=color2, below=of input, text width=4cm, text centered, minimum height=3.5em, yshift=-\shift cm-.1cm, xshift=-\shift cm+.2cm] {};
    }

  \node (quantum) [rectangle, draw, fill=color2, below=of input, text width=4cm, text centered, minimum height=3em, yshift=-0.3cm] {Quantum: VQAs \\ $|\Psi_{0}\rangle \xrightarrow{U(\bm{\theta})} |\Psi(\bm{\theta})\rangle \equiv \frac{\partial}{\partial \mathcal{\hat{O}}_k}$};

  \node (classical) [rectangle, draw, fill=color3, below=of quantum, text width=4cm, text centered, minimum height=3em] {Classical: L2O-$g^{\dagger}$ \\ $\arg \min_{\bm{\theta}\in \mathbb{R}^n} \mathcal{C(\bm{\theta})}$};

  \node (output) [rectangle, draw, fill=color1, below=of classical, text width=4cm, text centered, rounded corners, minimum height=3em] {Output};

  \node (expectation) [rectangle, draw, fill=color4, right=of classical, text width=2cm, text centered, minimum height=3em, xshift=.1cm, yshift=1cm] {Measure $\mathbb{E}[\{O_k\}]$};

  \node (update) [rectangle, draw, fill=color5, left=of classical, text width=2cm, text centered, minimum height=3em, xshift=-.1cm, yshift=1cm] {Update $\bm{\theta}$};

  \draw[-{Latex}] (input) -- (quantum);
  \draw[-{Latex}] (classical) -- (output);
  \draw[-{Latex}, rounded corners] (classical) -| (update);
  \draw[-{Latex}, rounded corners] (update) |- (quantum);
  \draw[-{Latex}, rounded corners] (quantum) -| (expectation);
  \draw[-{Latex}, rounded corners] (expectation) |- (classical);

  \begin{scope}[on background layer]
    \node (dashedbox) [rectangle, draw, dashed, fit=(quantum3) (expectation) (update) (classical), inner sep=0.3cm] {};
  \end{scope}

\end{tikzpicture}
\caption{\small
\textbf{Schematic Diagram for Variational Quantum Algorithms.} 
In classical optimization, we use \ours to find the optimal parameters \( \bm{\theta}^* \) for the given cost function \( \mathcal{C}(\bm{\theta}) \).
\ours iteratively updates the parameter \( \bm{\theta} \) until it converges to an approximate solution.
}
\label{fig:opt_diagram}
\end{figure}
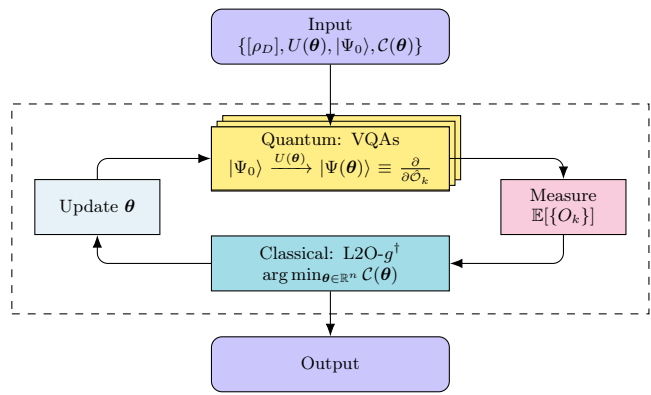

%% file: Tex/3_Theory.tex
\section{Theoretical Motivation}
\label{sec:theory}

In this section, we provide the theoretical motivation for \ours, a \textit{quantum-aware} learned optimizer that balances fast convergence and generalization. We first introduce the relevant mathematical formulation of the mirror descent framework. Extending the concept of optimization geometric of parameter space, we add the additional regularization term for penalized moving too far away from the direction of gradient to effectively balance convergence and generalization. This approach incorporates the geometry of the parameter space into the learned optimizer, differentiating it from previous works that used purely classical learned optimizers~\cite{verdon2019learning, wilson2019optimizing}. Specifically, we use a learnable vector that acts coordinate-wise on all the parameters  \( \bm{\theta} \), dynamically controlling convergence speed and generalization during the optimization process, adjusted by the types of VQA problems.

To accommodate the geometry of the optimization landscape, the mirror descent framework allows us to incorporate a generic distance function.
The update equation has the following form:
\begin{align}
\theta_{t+1} \leftarrow \arg \min_{\theta} \left\{ \left\langle \theta, \nabla \mathcal{C}(\theta_t) \right\rangle + D_{\Phi}\left( \theta, \theta_t \right) \right\}
\end{align}
where $ D_{\Phi}$ is the Bregman divergence to some function  \(\Phi : \mathbb{R} \rightarrow \mathbb{R}\) defined as:
\begin{align}
D_{\Phi}(\theta, \theta_t) = \Phi(\theta) - \Phi(\theta_t) - \langle \nabla \Phi(\theta_t), \theta - \theta_t \rangle.
\end{align}
For example, if we choose \( D_{\Phi}\left( \theta, \theta_t \right) = \frac{1}{2} \|\theta - \theta_t\|_2^2 \), it results in the gradient descent algorithm.
Alternatively, if we choose $D_{\Phi}(\theta, \theta_t) = \frac{1}{2} \|\theta - \theta_t\|_2^2 + \beta \sum_{i=1}^{t} \|\theta - \theta_i\|_2^2$ it corresponds to the momentum optimizer.
Finally, if we let \( D_{\Phi}\left( \theta, \theta + d\theta \right) = \mathrm{KL} \left( \theta \| \theta + d\theta \right) \), where \(\text{KL}\) represents the Kullback-Leibler divergence, it becomes the Quantum Natural Gradient~\cite{Stokes_2020} in the Mirror Descent framework:
\begin{align}
\theta_{t+1} \leftarrow \arg \min_{\theta} \left\{ \left\langle \bm{\theta}, \nabla \mathcal{C}(\theta_t) \right\rangle + \dfrac{1}{2\eta} \|u\|_{g(\theta_t)}^2 \right\}
\label{eq:md_qng}
\end{align}
where $\bm{u}=\left( \theta - \theta_t \right)$ and \( \|u\|_{g(\theta_t)}^2 \coloneq \left\langle u, g(\theta_t) u  \right\rangle \)
and \( g_{ij} \) is the Fubini-Study metric tensor define as:
\begin{align}
g_{ij} = \Re\left( \langle \partial_i \psi | \partial_j \psi \rangle - \langle \partial_i \psi | \psi \rangle \langle \psi | \partial_j \psi \rangle \right).
\label{eq:fubini}
\end{align}
In \Cref{eq:md_qng}, the first term is the linear approximation of \( \mathcal{C} \) at \( \bm{\theta}_t \) that is \( \mathcal{C}(\bm{\theta}) \approx \mathcal{C}(\bm{\theta}_t) + \langle \bm{\theta} - \bm{\theta}_t, \nabla \mathcal{C}(\bm{\theta}_t) \rangle \).
The second term $ \dfrac{1}{2\eta} \|u\|_{g(\theta_t)}^2 $ is the regularizer term penalized $\bm{\theta}$ moving too far away from $\bm{\theta}_t$ modified by metric tensor $g(\bm{\theta}_t)$.
We add an additional regularizer term as $\dfrac{\gamma_2}{2\eta} \|u-v\|_{g(\theta_t)}^2$ to penalized $\bm{\theta}$ moving too far away from $\bm{\theta}_t-\nabla \mathcal{C}(\theta_t)$ modified by metric tensor $g(\bm{\theta}_t)$.
We control the two regularizer terms with parameters $\gamma_1$ and $\gamma_2$.
With the addition of new regularizer terms, it provides the following update equation:
\begin{align}
\theta_{t+1} \leftarrow \arg \min_{\theta} \Bigg\{ 
\left\langle \theta, \nabla \mathcal{C}(\theta_t) \right\rangle 
& + \dfrac{\gamma_1}{2\eta} \|u\|_{g(\theta_t)}^2 \nonumber \\
& + \dfrac{\gamma_2}{2\eta} \|u-v\|_{g(\theta_t)}^2 \Bigg\}
\label{eq:md_two_reg}
\end{align}
where $\bm{v} = -\nabla \mathcal{C}(\theta_t)$, and $\gamma_1>0, \gamma_2>0$ are parameters that modulate between two regularizers.
The modified gradient given by the optimization problem in \Cref{eq:md_two_reg} is
\begin{equation}
\nabla \mathcal{\widetilde{C}}(\theta) = \left( \frac{\gamma_1}{\gamma_1-\gamma_2}g^{\dagger} - \frac{\gamma_2}{\gamma_1-\gamma_2}I \right) \bm{v}.
\label{eq:modify_g}
\end{equation}

%% file: Tex/2_Methodology.tex
\section{Methodology}
\label{sec:method}

\subsection{Background and Problem Formulation}
A parameterized quantum circuit \( U(\bm{\theta}) \) has parameters \( \bm{\theta} \in \Theta \subseteq \mathbb{R}^N \). 
To enhance the expressibility of VQAs, \( U(\bm{\theta}) \) usually consists of \( l \) layers of unitaries:
\begin{equation}
    U(\bm{\theta}) = U_l(\bm{\theta}_l)\cdots U_2(\bm{\theta}_2)U_1(\bm{\theta}_1),
\end{equation}
where $\bm{\theta}_l$ is the $l$-th layers of $\bm{\theta}$.
Each unitary can be written as
\begin{equation}
    U_l(\bm{\theta}_l) = \prod_m e^{-i\theta_m \hat{H}_m}W_m
\end{equation}
where $H_m$ is Hermitian operator and $W_m$ is an unparameterized unitary.
For example, QAOA with $p$ layers has the following unitary form $U_l(\bm{\alpha}_l, \bm{\beta}_l) = e^{-i\beta_l \hat{H}_M} e^{-i\gamma_l \hat{H}_C}$, where $\hat{H}_C$ is the cost Hamiltonian, $\hat{H}_M$ is the mixer Hamiltonian, and the respective parameters are $\bm{\gamma} \in \mathbb{R}^p$ and $\bm{\beta} \in \mathbb{R}^p$.
The cost function encode by VQAs is some observable $\hat{H}$ in terms of parameters $\bm{\theta}$ define as: 
\begin{align}
    \mathcal{C}(\bm{\theta}) = \langle \psi_0 | U^\dagger(\bm{\theta}) \hat{H} U(\bm{\theta}) | \psi_0 \rangle,
\end{align}
with the initial state $\ket{\psi_0}$ on $N_q$ qubits.
Parameter usually initialize with ground state $\ket{0}^{\otimes N_q}$ or other heuristic strategies~\cite{Lee_2021, Jain_2022}.
Given sample $\rho_d$ from dataset of training state $[\rho_D]$. 
The objective for VQAs problem is finding the optimal parameters $\bm{\theta^*}$ such that
\begin{equation}
    \bm{\theta^{*}} = \arg \min_{\theta\in \mathbb{R}^n} \sum_{\rho_d \in [\rho_D]} \mathcal{C}_{\rho_d}(\bm{\theta}).
\end{equation}
The overview diagram for VQAs is illustrated in \Cref{fig:opt_diagram}.
VQA is a hybrid framework consisting of a parameterized quantum circuit and leverages a classical computer to iteratively update the parameters $\bm{\theta}$ until it converges.
However, training a PQC with a classical optimizer to find the optimal parameters $\bm{\theta}^*$ is generally an $\mathsf{NP}$-hard problem~\cite{PhysRevLett.127.120502}.
A common-use algorithm is the descent algorithm which can be written as a general form
\begin{equation}
    x_{k+1} = x_{k} + t_{k}\Delta x_{k}
    \label{eq:decent_alg}
\end{equation}
which produce a optimization trajectory $[x_{k}]$.
A natural choice is setting $\Delta x = -\nabla f(x)$, and it leads to the gradient descent algorithm.
However, hand-designed optimizers critically depend on the choice of hyperparameters.
To tackle this issue, learning to optimize (L2O) has provided a new paradigm by training a small neural network to "learn" the update rule, leveraging the neural network's capacity as a universal function approximator~\cite{hornik1989multilayer}.
Another major challenge for the trainability of VQAs is barren plateaus and narrow gorges in PQCs cost function landscape~\cite{mcclean2018barren, arrasmith2022equivalence}.
The variance of gradient respect with any $\theta_\mu \in \bm{\theta}$ vanishes exponentially with the number of qubits $N_q$:
\begin{equation}
    \text{Var}_{\bm{\theta}} [ \partial_{\mu} \mathcal{C}](\bm{\theta}) \le F(N_q) 
\end{equation}
with $F(N_q) \in \mathcal{O}\left( b^{-N_q} \right)$ for some constant $b > 1$.

In \ours, we replace $t_k\Delta x_{k}$ with a \textit{quantum-aware} Long Short-Term Memory Model denote as L2O-$g^{\dagger}$.
$\text{L2O-}g^{\dagger}(\cdot; \bm{\phi})$ is a parametric update function with meta-parameters $\bm{\phi} \in \Phi$ which takes the input state $\bm{z}_t$ then update PQC parameter to $\bm{\theta}_{t+1}$.
The corresponding update equation analogy to \Cref{eq:decent_alg} is given by the following parameter update rule
\begin{equation}
    \bm{\theta}_{t+1} = \bm{\theta}_t - \text{L2O-}g^{\dagger}(\bm{z}_t; \bm{\phi}).
    \label{eq:ours_update}
\end{equation}
The goal of VQAs is to minimize the expected loss for cost function $\mathcal{C}$ at $k$-step which we define as $L_k(\bm{\theta}) = \mathbb{E}_{\rho_d} [\mathcal{C}_{\rho_d}(\bm{\theta}_k)]$.
The objective for \ours is find optimal $\bm{\phi}^{*}$ that minimize the outer-loss $ \mathcal{L}(\bm{\phi}; T)$,
\begin{equation}
\begin{aligned}
    \mathcal{L}(\bm{\phi}; T) = \sum_{t=1}^{T} w_t L_t(\bm{\theta}_t), \quad
    \bm{\phi}^{*} = \arg \min_{\bm{\phi} \in \Phi} \mathcal{L}(\bm{\phi}; T)
\end{aligned}
\end{equation}
where $w_t$ is weighted term and $T$ is the horizon for trajectory.

\subsection{\ours Architecture}
\label{sec:l2o_arch}

In this section, based on our theoretical motivation in \Cref{sec:theory}.
We propose \ours, a \textit{quantum-aware} learned optimizer that dynamically balances between convergence speed and generalization.
In \Cref{fig:loss_traj} gives a toy example of \ours on optimizing the MaxCut problem with QAOA. 
The optimization trajectory learned by \ours exhibits totally different strategy compared with any first-order and second-order optimizer.
It dynamically adjusts step size and converges to a minimum with only a few steps.

\begin{figure*}[t]
    \centering
    \vspace{-1em}
    \includegraphics[scale=0.20]{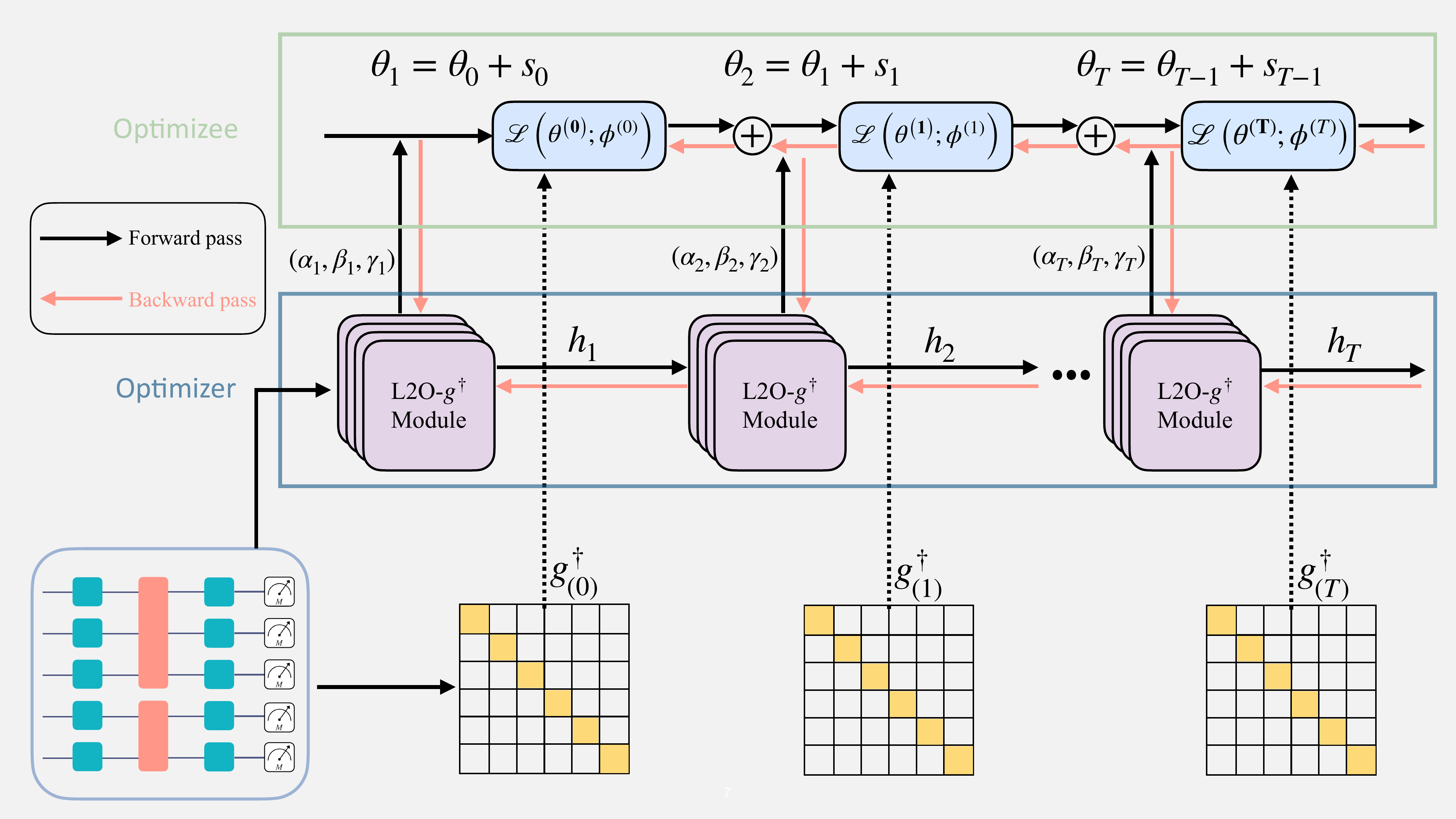}
    \vspace{-1em}
    \caption{\small
    \textbf{Overall Architecture of Quantum-Aware Learned Optimizer \ours.}
    Given the parameters \(\bm{\theta}_t\) from PQC, \ours dynamically balances between distribution space optimization and parameter space optimization using the learning strategy learned by the \ours module.
    The \ours module consists of a recurrent neural network, and the output values pass through a linear layer, returning three vectors \( (\bm{\alpha}_t, \bm{\beta}_t,\bm{\gamma}_t) \) at the time step \(t\) in the optimizee trajectory \([x_T]\).
    The output for distribution space optimization is modified by the Fubini-Study metric tensor \( g^{\dagger} \). Along with the update step $\bm{\eta_t} = \exp \left[ \lambda_b \bm{\alpha}_t \right]\in\mathbb{R}^{N}$ and the update direction $\bm{v}=\lambda_a \bm{\beta}_{t}\in\mathbb{R}^{N}$, these update the parameters to \(\bm{\theta}_{t+1}\).
    In the forward pass, the \ours module takes the hidden state \(\bm{h}_t\), cell state \( \bm{c}_t \), and pre-processed input state \( \bm{z}_t \).
    The backward pass happens after every $T_{\text{unrolled}}^{(i)}$ at the \( i \)-stage of curriculum learning, where it gradually increases to facilitate exploration and exploitation.
    }
    \label{fig:l2o-g}
\end{figure*}

We describe the overall architecture of our quantum-aware learned optimizer in this section. 
\Cref{fig:l2o-g} illustrates the architecture of our proposed method. It consists of a learned optimizer, a recurrent neural network with parameters $\bm{\phi} \in \Phi$. 
The input state $\bm{z}_{t}=(\log(\lvert\nabla\rvert), \mathrm{sgn}(\nabla))$ follows the method by \citet{andrychowicz2016learning}. 
It returns the parameter update for the optimizee, a parameterized quantum circuit.
We summarize the entire \ours as a function:
\begin{align}
\bm{\alpha}_t, \bm{\beta}_t, \bm{\gamma}_t, \bm{h}_t, \bm{c}_t = \text{\ours}(\bm{z_t}, \bm{h}_{t-1}, \bm{c}_{t-1})
\end{align}
where $\bm{h}_t$ and $\bm{c}_t$ are the hidden state and cell state of the Long Short-Term Memory model at time step $t$. 
It outputs the update step $\bm{\eta_t} = \exp \left[ \lambda_b \bm{\alpha}_t \right]\in\mathbb{R}^{N}$ and the update direction $\bm{v}=\lambda_a \bm{\beta}_{t}\in\mathbb{R}^{N}$ which act coordinate-wise on each parameter $\bm{\theta}$ in PQC. 
The scale factor are $\lambda_a=\lambda_b=0.01$.

Inspired by \Cref{eq:modify_g}, we replace \(\gamma_1\) and \(\gamma_2\) with a coordinate-wise learnable vector \(\bm{\gamma_t} \in (0,1)^N\), defined as \(\bm{\gamma_t} = \text{Sigmoid}(\text{Linear}(h_t))\).
It modulates the balance between two terms expressed as 
$\nabla \mathcal{\widetilde{C}} \left(\bm{\theta}_{t} \right) = \left( \left( 1- \bm{\gamma}_t \right) g^{\dagger}_t + \bm{\gamma}_t I \right) \bm{v}$
where $\bm{B}_t=\left( \left( 1- \bm{\gamma}_t \right) g^{\dagger}_t + \bm{\gamma}_t I \right)$ acts as a effective preconditioner. 
It combines the update direction modified by the parameter space and the distribution space controlled by the learnable vector $\bm{\gamma}$ that auto-adjusts during the optimization process. Combining all, it gives us the updated equation for \ours:
\begin{align}
\bm{\theta}_{t+1}=\bm{\theta}_{t} -  \bm{\eta}_t \circ \bm{B}_t \bm{v}_t 
\end{align}
where $\bm{B}_t$ is a diagonal matrix.

\subsection{Training Details}
\label{sec:train_detail}

Learned optimizers are known to be unstable and difficult to train \cite{metz2019understanding}. 
Backpropagation through unrolled optimization can result in a strong bias for short truncations or exploding norms for long truncations.
Therefore, we adopt the curriculum learning technique for the learned optimizer proposed by \citet{chen2020training}. 
It gradually increases the difficulty of the problem to facilitate faster convergence, better generalization, and stable training. 
In \ours, the unrolled length gradually increases in this sequence: $T_{\text{unrolled}}=[10, 20, 40, 60, 80, \ldots]$. 
When training steps are at the $i^{\text{th}}$ stage, the validation length mismatches at the $(i+1)^{\text{th}}$ stage. 
Training terminates when none of the validation losses at the $(N+1)^{\text{th}}$ stage is lower than the $N^{\text{th}}$ stage-trained model.

In the following experiment detailed in \Cref{sec:exp}.
We train two models, each using a single PQC as training data: a generic PQC for solving VQE and QML tasks, and a QAOA circuit for MaxCut and Sherrington-Kirkpatrick Model tasks.

\subsection{Baseline Optimizers}
\label{sec:base_opt}

There is a collection of hand-designed optimizers \cite{schmidt2021descending} for optimizing high-dimensional and non-convex deep learning models.
In our experiments, we select several popular optimizers, including RMSprop~\cite{hinton2012rmsprop}, Adam~\cite{kingma2017adam}, Gradient Descent, Quantum Natural Gradient Descent~\cite{Stokes_2020}, Momentum~\cite{rumelhart1986learning}, and Adagrad~\cite{duchi2011adaptive}.

%% file: Tex/5_Experiments.tex
\section{Experiments}
\label{sec:exp}

In this section, we benchmark \ours on a broad spectrum of VQA problems.
Our experiments can organized into the following three parts, respectively showing:
\begin{itemize}
    \item \ours can optimize general parameterized quantum circuits by training on a single instance, and it surpasses the hand-design optimizer in terms of convergence speed.
    \item By training \ours on a single general PQC, it can tackle all kinds of real VQAs problems including VQE for chemistry, QAOA for MaxCut and Sherrington-Kirkpatrick Model, and QML for data-reuploading circuits (see \Cref{sec:rand_pqc,sec:vqe,sec:maxcut,sec:sk_model,sec:reupload}). Additionally, out-of-box \ours can match or outperform hyperparameter tuned optimizer.
    \item Conducting ablation study on various parts of \ours, and show that it surpasses previous learned optimizer baselines for optimizing PQC on generalizability and convergence speed (see \Cref{sec:ablation}).
\end{itemize}

\paragraph{Tasks Setting.}
The tasks setting overview are the following:
\begin{itemize}
    \item Random Parameterized Quantum Circuit (see \Cref{sec:rand_pqc}): \citet{farhi2014quantum} used it to showcase the effect of the barren plateau. 
    We follow the same experimental setup in \citet{Stokes_2020} for the baseline optimizer.
    \item VQE~\cite{Peruzzo_2014} for Chemistry (see \Cref{sec:vqe}): We implement various ansatz including Hardware Efficient Ansatz (HEA) and $U^{(1)}_{\text{ent}}$ ansatz~\cite{Barkoutsos_2018}, Unitary Coupled-Cluster Singles and Doubles (UCCSD) ansatz~\cite{Barkoutsos_2018} and calculate various ground states of molecules including $\mathrm{H}_2$, $\mathrm{H}_3^+$, $\mathrm{H}_4$, $\mathrm{LiH}$, $\mathrm{BeH}_2$, and $\mathrm{H}_2\mathrm{O}$.
    \item QAOA~\cite{farhi2014quantum} for MaxCut (see \Cref{sec:maxcut}): We calculate the approximation ratio by QAOA for the MaxCut problem with randomly generated Erdős-Rényi graphs and compare it with learning rate grid search quantum natural gradient descent algorithm~\cite{Stokes_2020}.
    \item QAOA~\cite{farhi2014quantum} for Sherrington-Kirkpatrick Model (see \Cref{sec:sk_model}): We compare the final objective value for a wide variety of optimizers for optimizing the Sherrington-Kirkpatrick model with QAOA of various layers $p_{\text{layer}}$.
    \item QML~\cite{Biamonte_2017} for Data Re-upload Circuit~\cite{P_rez_Salinas_2020} (see \Cref{sec:reupload}): We implement the same experimental setup from \citet{P_rez_Salinas_2020} and compare a single trial \ours with second-order optimizer L-BFGS-B and learning rate tuned Adam and Momentum.
\end{itemize}

\subsection{Random Parameterized Quantum Circuit}
\label{sec:rand_pqc}

\paragraph{Setup.}
We employ the circuit design in \citet{mcclean2018barren}. 
It consists of: Initialized with $\ket{\psi_1} = (R_Y(\frac{\pi}{4})^{\otimes N_q}) \ket{0}^{\otimes N_q}$, then followed by $l$ layers given by $\ket{\psi_l} = (U_{\text{CZ}}^{(l)} U_{\text{P}}^{(l)}) \cdots (U_{\text{CZ}}^{(1)} U_{\text{P}}^{(1)}) \ket{\psi_1}$. 
In particular, $U_{\text{P}}^{(i)} = \bigotimes_{j=1}^{n} R_{P_{j}}(\theta_{i,j})$ and $U_{\text{CZ}}^{(i)} = \prod_{j=1}^{n-1} \text{CZ}_{j, j+1}$, where $P_j \in \{X, Y, Z\}$, and $\theta_{i,j} \in [0, 2\pi)$.

The objective is a single Pauli \(ZZ\) operator on the first and second qubits, \(\hat{\sigma}_z \otimes \hat{\sigma}_z\).
The circuit is visualized in \Cref{fig:random_circuit1}. 
The hyperparameter settings follow \citet{Stokes_2020} when possible. 
The default settings are organized in \Cref{app:hyperparameter}. 
In this experiment, \ours is trained on a single PQC with the configuration of $(N_q,l)=(7,5)$. 
By controlling the number of qubits $N_q$ and layers $l$, we can demonstrate \ours robustness against different numbers of optimization dimensions beyond its training PQC.
\input{Tex/random_circuit}

\begin{figure*}[htb!]
    \centering
    \begin{subfigure}[b]{0.32\textwidth}
        \includegraphics[width=\linewidth]{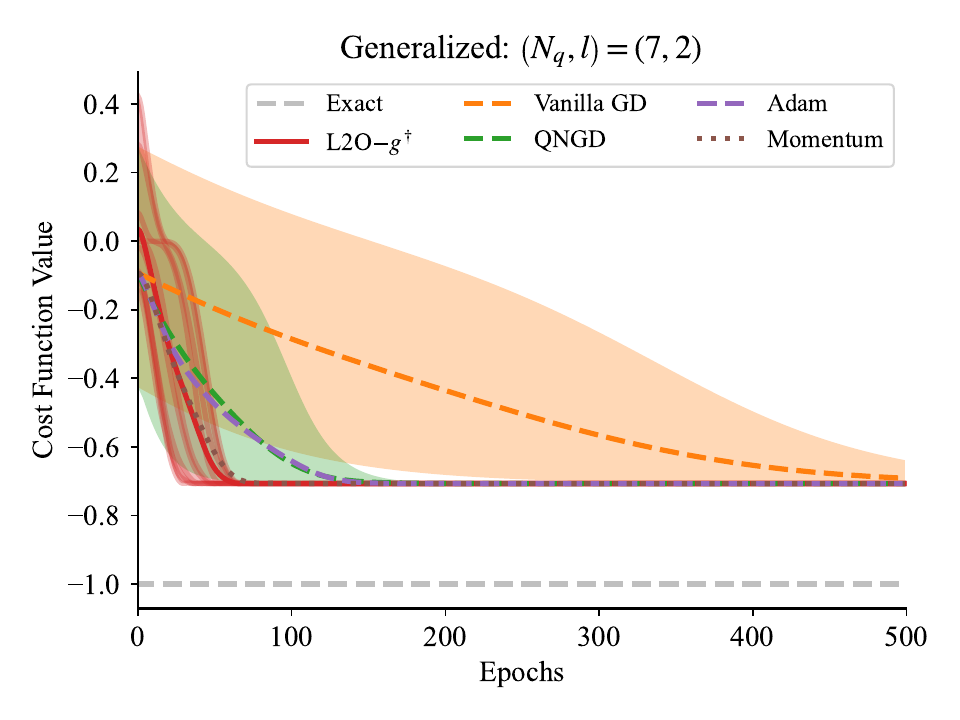}
        \caption{$N_q=7, l=2$}
    \end{subfigure}
    \hfill
    \begin{subfigure}[b]{0.32\textwidth}
        \includegraphics[width=\linewidth]{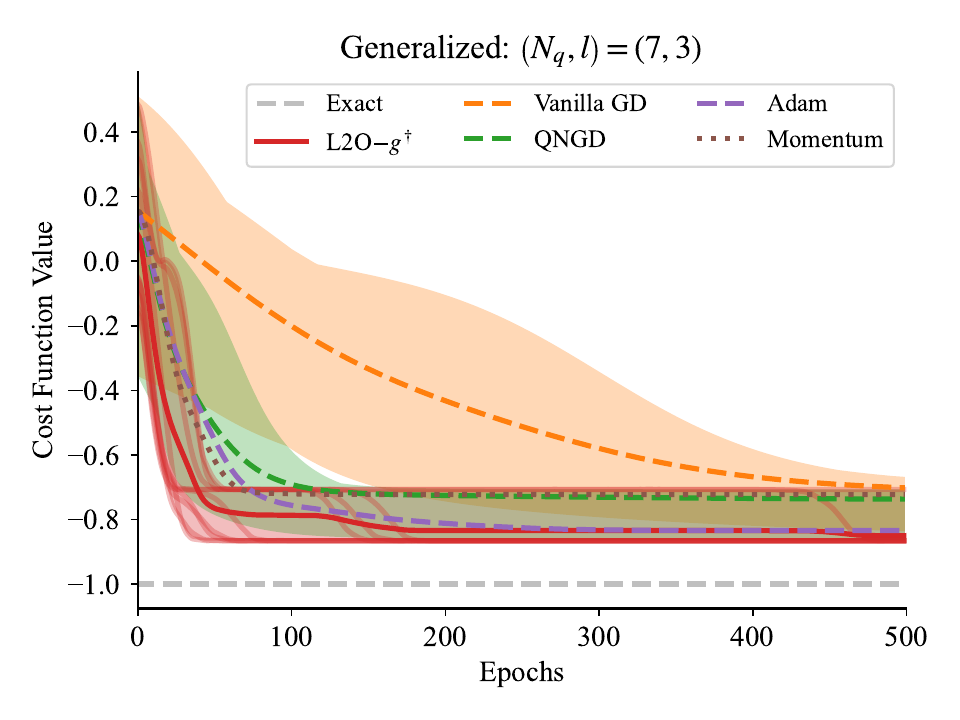}
        \caption{$N_q=7, l=3$}
    \end{subfigure}
    \hfill
    \begin{subfigure}[b]{0.32\textwidth}
        \includegraphics[width=\linewidth]{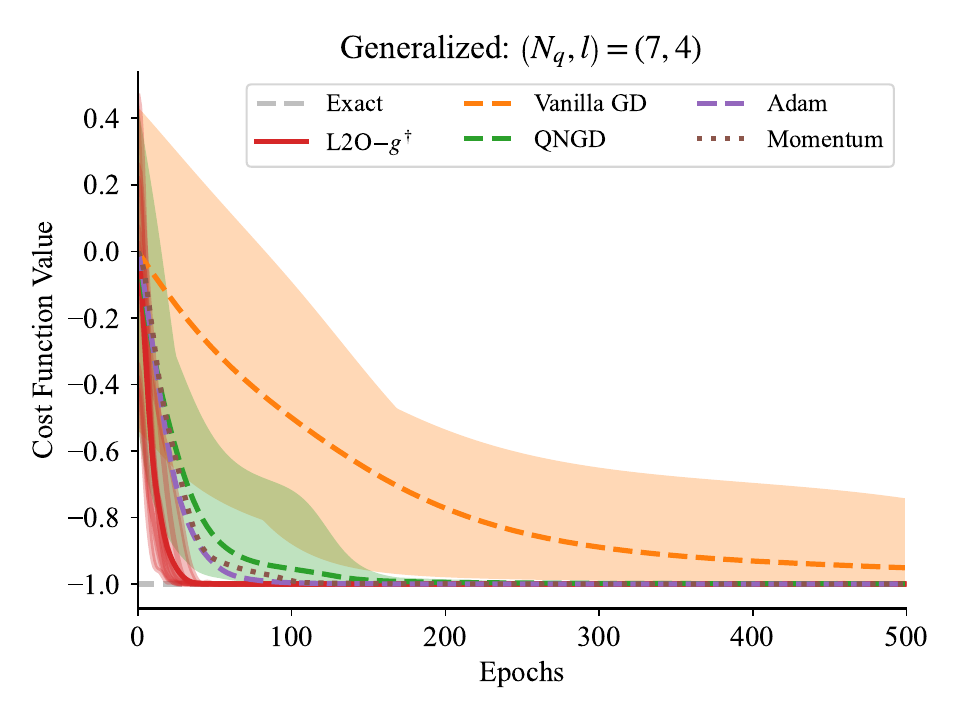}
        \caption{$N_q=7, l=4$}
    \end{subfigure}

    \vspace{1em} %
    \begin{subfigure}[b]{0.32\textwidth}
        \includegraphics[width=\linewidth]{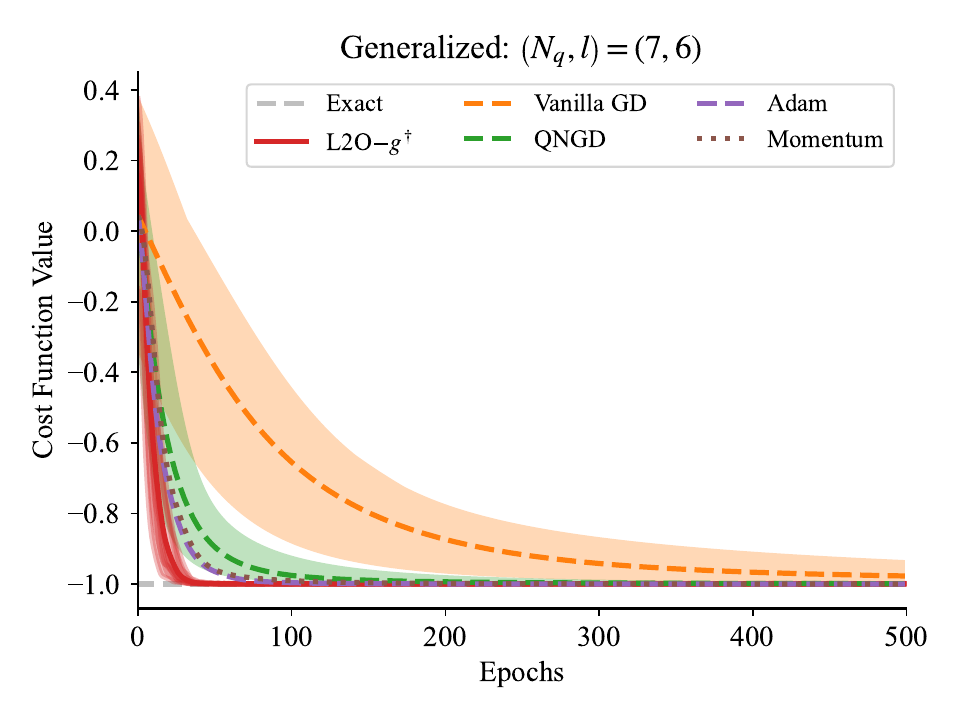}
        \caption{$N_q=7, l=6$}
    \end{subfigure}
    \hfill
    \begin{subfigure}[b]{0.32\textwidth}
        \includegraphics[width=\linewidth]{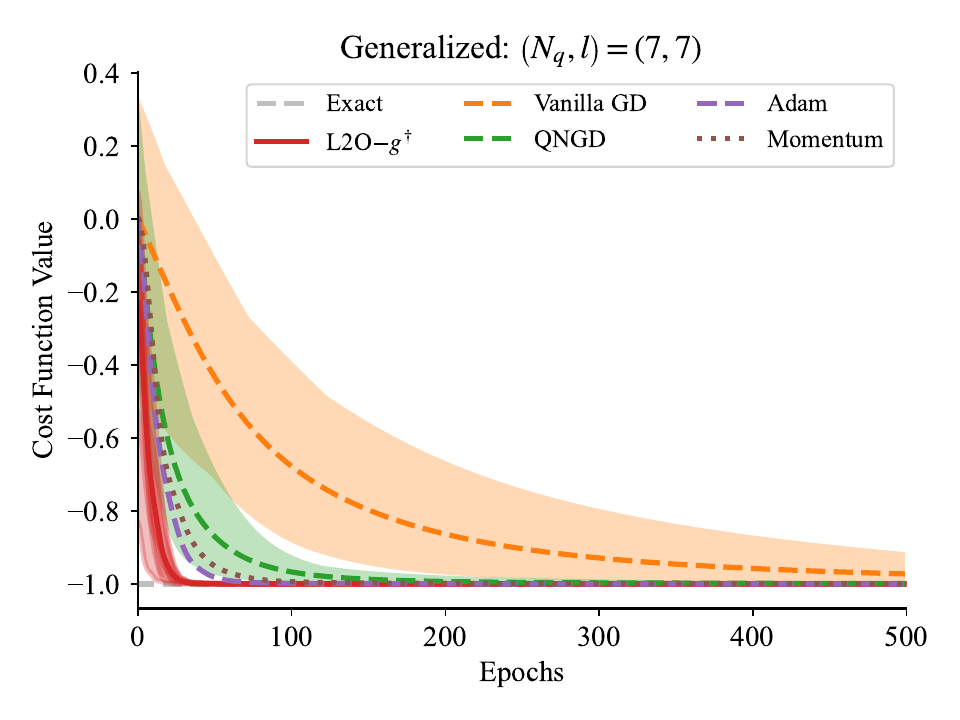}
        \caption{$N_q=7, l=7$}
    \end{subfigure}
    \hfill
    \begin{subfigure}[b]{0.32\textwidth}
        \includegraphics[width=\linewidth]{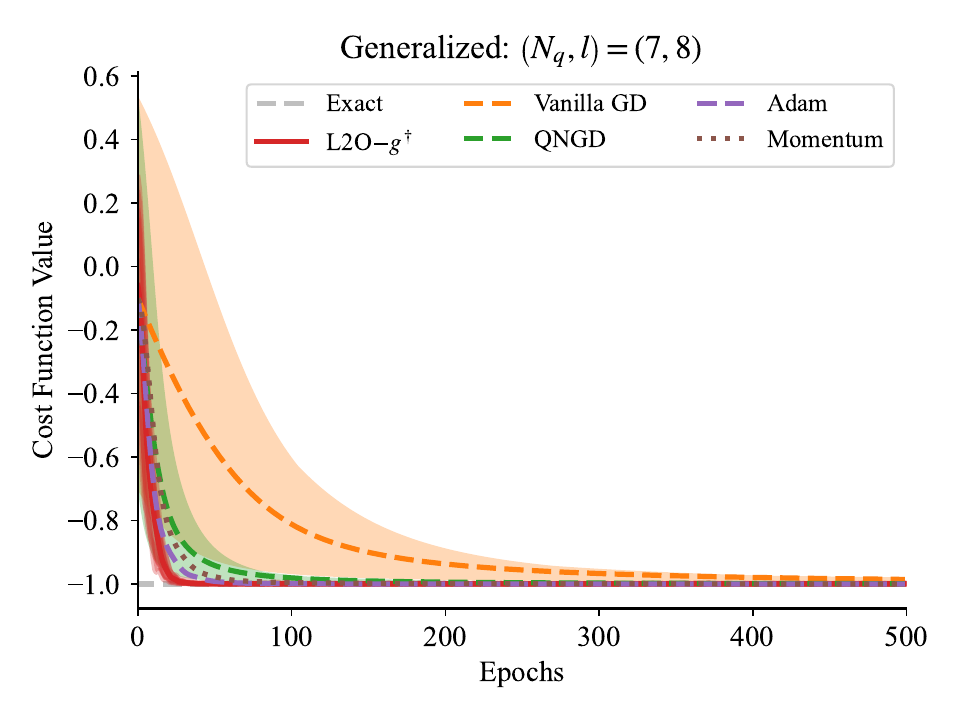}
        \caption{$N_q=7, l=8$}
    \end{subfigure}

    \vspace{1em} %
    \begin{subfigure}[b]{0.32\textwidth}
        \includegraphics[width=\linewidth]{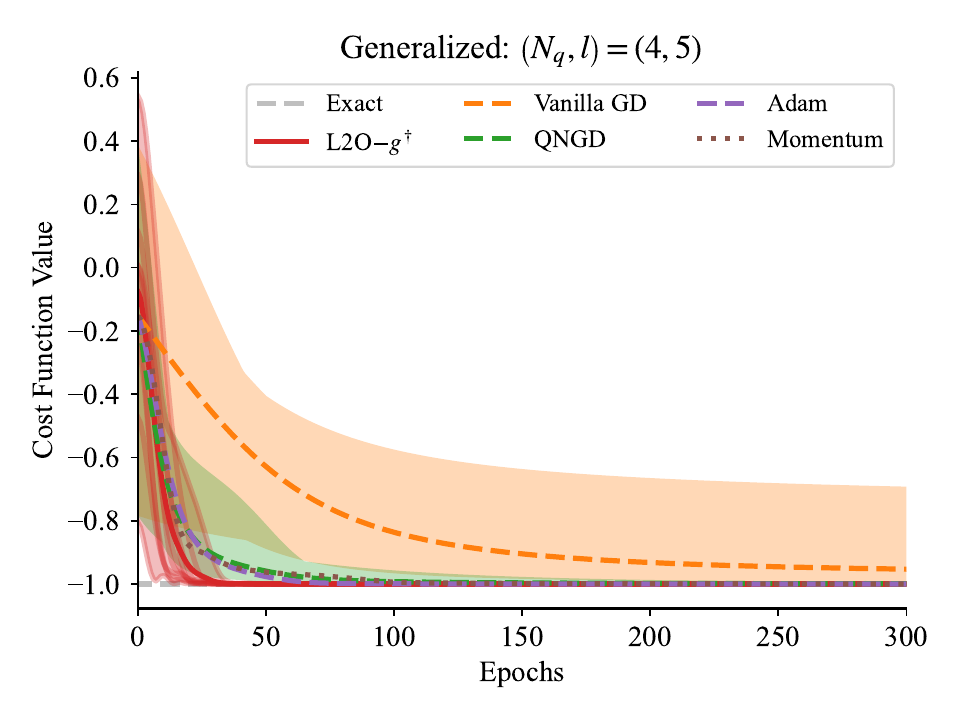}
        \caption{$N_q=4, l=5$}
    \end{subfigure}
    \hfill
    \begin{subfigure}[b]{0.32\textwidth}
        \includegraphics[width=\linewidth]{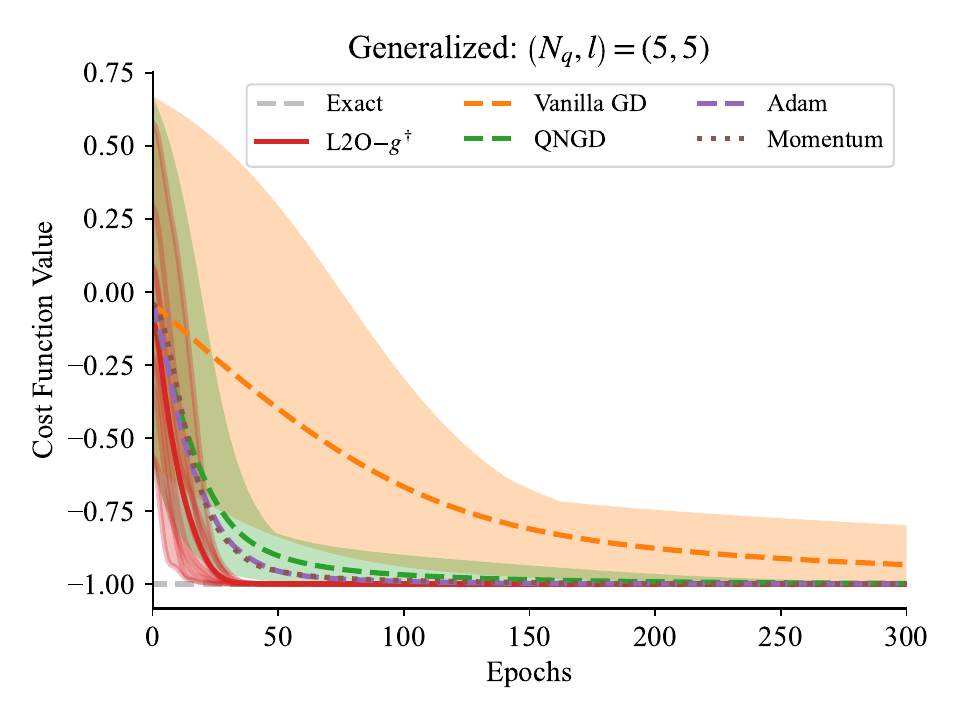}
        \caption{$N_q=5, l=5$}
    \end{subfigure}
    \hfill
    \begin{subfigure}[b]{0.32\textwidth}
        \includegraphics[width=\linewidth]{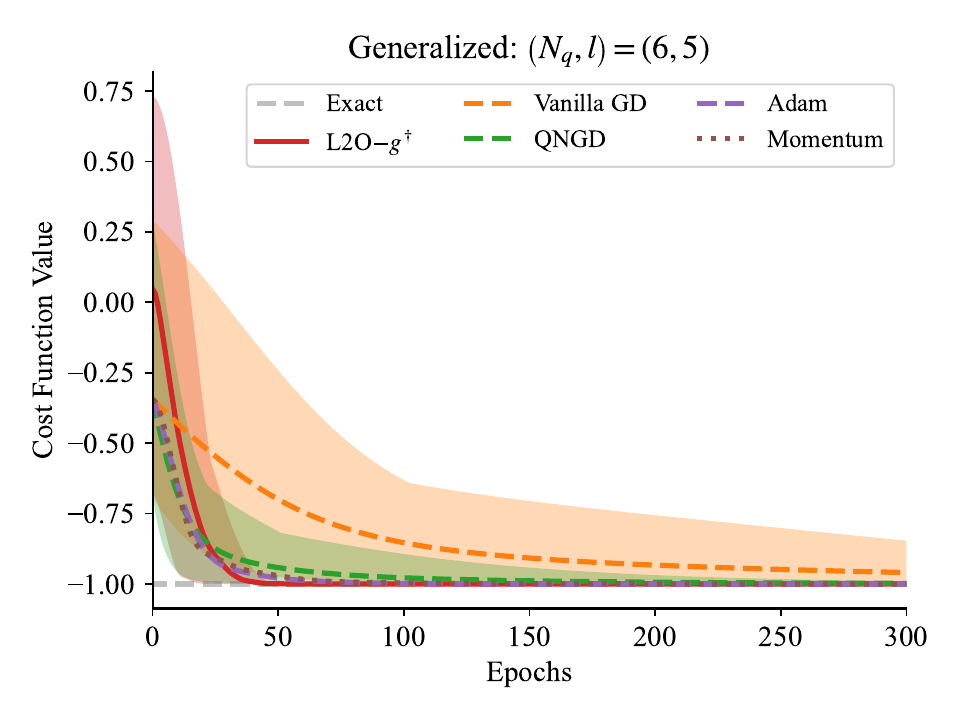}
        \caption{$N_q=6, l=5$}
    \end{subfigure}

    \vspace{1em} %
    \begin{subfigure}[b]{0.32\textwidth}
        \includegraphics[width=\linewidth]{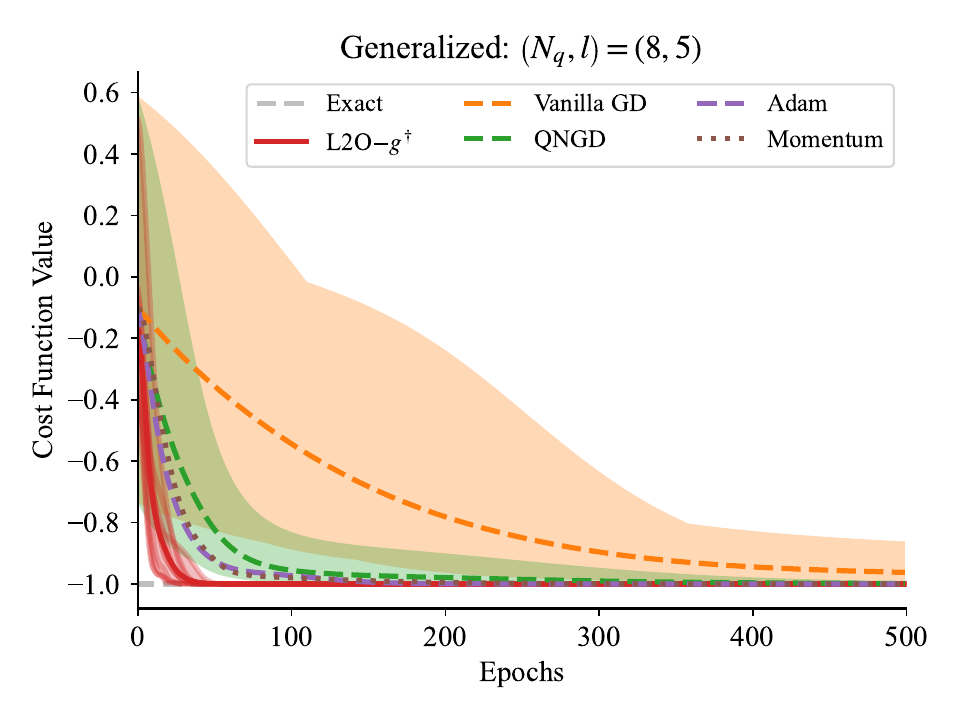}
        \caption{$N_q=8, l=5$}
    \end{subfigure}
    \hfill
    \begin{subfigure}[b]{0.32\textwidth}
        \includegraphics[width=\linewidth]{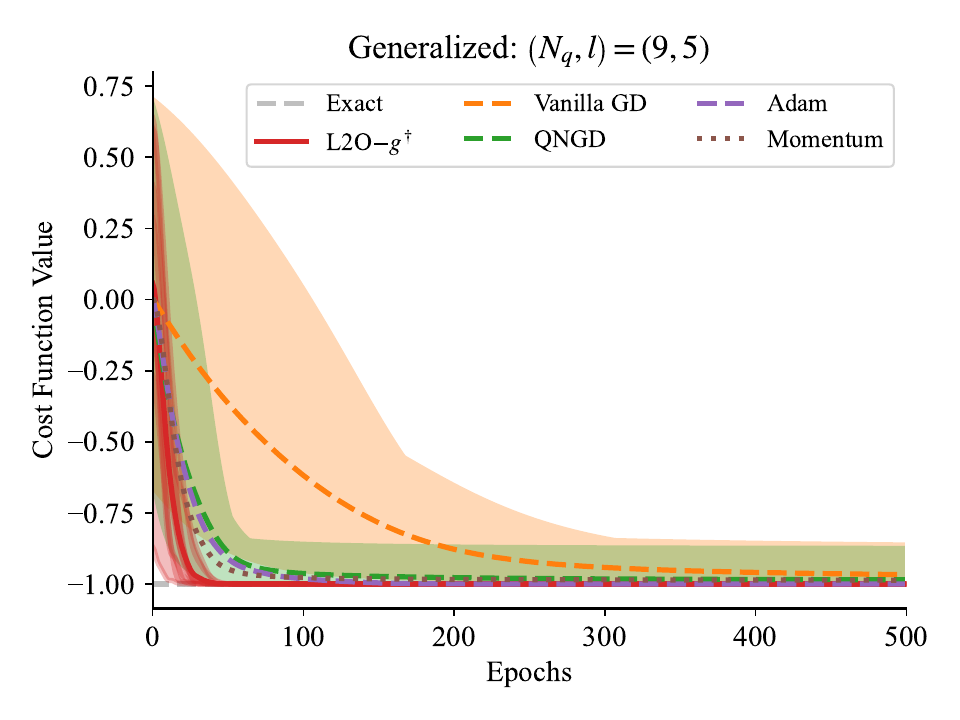}
        \caption{$N_q=9, l=5$}
    \end{subfigure}
    \hfill
    \begin{subfigure}[b]{0.32\textwidth}
        \includegraphics[width=\linewidth]{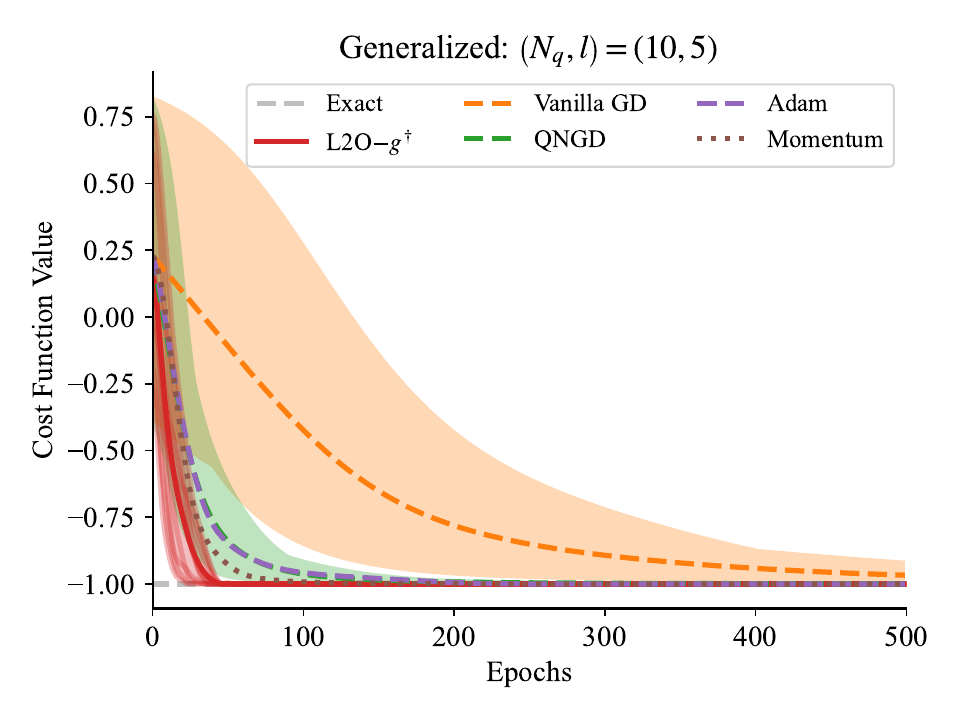}
        \caption{$N_q=10, l=5$}
    \end{subfigure}

    \caption{\small
    \textbf{\ours Generalized to Different Random PQC.} The upper two rows from (a) to (f) show \ours abilities to adapt to different numbers of layers $l$ from $2$ to $8$ with $N_q=7$. In the case of $l=2$ and $l=3$, all methods do not converge to the optimal value of $-1$ due to a lack of tunable parameters. However, \ours still reaches the minimum value faster than other baseline optimizers. The bottom two rows from (g) to (l) show \ours abilities to adapt to different numbers of qubits $N_q$ from $4$ to $10$ with $l=5$. \ours consistently outperforms in convergence speed on all configurations and is less sensitive to the initial parameter. In all experiments, we average over 10 different sequences of single qubit gates and plot the average (full color) and each loss curve (lighter color), along with the baseline average (solid line) and standard deviation (ribbon).}
    \label{fig:combined_random_circuits}
\end{figure*}

\paragraph{Results.} We summarize our results for optimizing Random PQC in \Cref{fig:combined_random_circuits} with various qubits $N_q$ and layers $l$.
\ours demonstrates robustness against entirely different random circuits and varying numbers of parameters compared to its training instances.
With only one random PQC in its training stage, this highlights \ours ability to generalize \textit{optimization strategy}.

\subsection{Variational Quantum Eigensolver (VQE)}
\label{sec:vqe}

\paragraph{Backgrounds.}
Variational quantum eigensolver (VQE) was originally proposed by \citet{Peruzzo_2014} and has become an important algorithm in the VQAs family. 
It is based on the variational principle of quantum mechanics:
\begin{equation}
E_0 \leq \langle \psi | \hat{H} | \psi \rangle.
\end{equation}
Given a trial wavefunction and Hamiltonian \(\hat{H}\) that encodes the problem, VQE finds the ground state energy of the system \(E_0\) with optimal parameters \(\bm{\theta}^*\):
\begin{equation}
\lambda_{\min} = E_0 \approx \bra{\psi(\bm{\theta}^*) } \hat{H}  \ket{\psi(\bm{\theta}^*)}.
\label{eq:vqe_ground}
\end{equation}
The Hamiltonian in the second quantization form for a many-body system can be written in terms of the creation operator $\bm{c}^\dagger$ and annihilation operator $\bm{c}$:
\begin{equation}
\hat{\bm{H}} = \sum_{p,q} h_{pq} \bm{c}_p^\dagger \bm{c}_q + \frac{1}{2} \sum_{p,q,r,s} h_{pqrs} \bm{c}_p^\dagger \bm{c}_q^\dagger \bm{c}_r \bm{c}_s.
\end{equation}

\paragraph{Ansatz.}
The simplest ansatz selection is hardware-efficient ansatz (HEA) which is a parameterized quantum circuit tailored to quantum devices~\cite{kandala2017hardware, o2016scalable, hu2022benchmarking, sun2024toward}. 
It works flexibly but it needs to span a large Hilbert space so at worst it scales exponentially in the worst case.
HEA can be represented in the following form 
\begin{equation}
    \prod_{l=1}^{L} \left( \prod_{i=1}^{n} U(\theta_{i, l}) \prod_{(i,j) \in E} \text{CNOT}_{ij} \right)
\end{equation}
where $n$ is the number of qubits, $L$ is the number of layers of gates, $U(\theta_{i, l})$ represents a parameterized single-qubit gate on the $i$-th qubit in the $l$-th layer, and $E$ represents the set of qubit pairs $(i, j)$ connected by CNOT gates within the hardware's connectivity graph.

The second and third type of ansatz proposed by \citet{Barkoutsos_2018} are the $U^{(1)}_{\text{ent}}$ ansatz and Unitary Coupled-Cluster Singles and Doubles (UCCSD) ansatz.
$U^{(1)}_{\text{ent}}$ adjust HEA with the constraint of preserving a constant number of particles in the following form proposed by~\cite{arrazola2022universal, Barkoutsos_2018}
\begin{equation}
    |\Psi(\bm{\theta})\rangle = U_D(\bm{\theta}) U_{\text{ex}} \ldots U_1(\bm{\theta}) U_{\text{ex}} U_0(\bm{\theta}) |\Phi_0\rangle
\end{equation}
where it consist of single-qubit rotations $U_i(\bm{\theta})$ at $i$ layers and entagling \textit{drift} operation $U_{\text{ex}}$.
In particular $U_{\text{ex}}$ can be particle conserving two-parameter exchange-type gate~\cite{Barkoutsos_2018} with the following form in \Cref{eq:part_ex}.
\begin{equation}
U_{\text{ex}}(\theta_1, \theta_2) = \begin{pmatrix}
1 & 0 & 0 & 0 \\
0 & \cos\theta_1 & e^{i\theta_2} \sin\theta_1 & 0 \\
0 & e^{-i\theta_2} \sin\theta_1 & -\cos\theta_1 & 0 \\
0 & 0 & 0 & 1 
\end{pmatrix}
\label{eq:part_ex}
\end{equation}

The third type of ansatz we implement in our experiments is the Unitary Coupled-Cluster Singles and Doubles (UCCSD) ansatz~\cite{Barkoutsos_2018}. 
The Coupled Cluster (CC) ansatz uses these excitations to create a wave function from the Hartree-Fock state as \(\Psi = e^{\tilde{\bm{T}}} \Phi_{\text{HF}}\).
\begin{equation}
e^{\bm{T} - \bm{T}^\dagger} \quad \text{where} \quad \bm{T} = \bm{T}_1 + \bm{T}_2
\end{equation}
\begin{equation}
\bm{T}_1 = \sum_{i,a} t_i^a \bm{a}_a^\dagger \bm{a}_i 
,\quad 
\bm{T}_2 = \sum_{i,j,a,b} t_{ij}^{ab} \bm{a}_a^\dagger \bm{a}_b^\dagger \bm{a}_i \bm{a}_j
\end{equation}
$\bm{T}_1$ represents single excitations, with $t_i^a$ denoting the amplitude for exciting an electron from the occupied spin-orbital $i$ to the virtual spin-orbital $a$. 
$\bm{T}_2$ represents double excitations, with $t_{ij}^{ab}$ as the amplitude for exciting electrons from the occupied orbitals $i$ and $j$ to the virtual orbitals $a$ and $b$. 
Here, $\bm{a}_i^\dagger$ and $\bm{a}_i$ are fermionic creation and annihilation operators, respectively.

\paragraph{Setup.}
We implement three ansatz: HEA, $U^{(1)}_{\text{ent}}$, and UCCSD. 
For HEA, we show that by training \ours on a single bond length of $\mathrm{H}_2$, it can generalize to different bond lengths with relatively small errors. 
For $U^{(1)}_{\text{ent}}$ ansatz, we choose $\mathrm{LiH}$, $\mathrm{BeH}_2$, and $\mathrm{H}_2\mathrm{O}$ molecules and calculate the ground state energy with various repeated entangle blocks in the ansatz, ranging from $D=1$ to $D=30$ entangle blocks. 
For UCCSD ansatz, we choose $\mathrm{H}_2$, $\mathrm{H}_3^+$, and $\mathrm{H}_4$ molecules and calculate the approximate ground state energy for various inter-atomic radii.
We report the mean of ten runs and the $90\%$ confidence interval, along with two baseline optimizers: RMSprop and Adam. 

All the optimizations for various ansatz are taken from randomly initialized parameters and the objective values after 200 iterations.
We use randomly initialized parameters for the HEA and UCCSD ansatz and initialize with the Hartree-Fock state for the \( U^{(1)}_{\text{ent}} \) ansatz. 
The molecular data are provided by Pennylane datasets~\cite{Utkarsh2023Chemistry}.

\begin{figure*}[htb!]
    \centering
    \begin{minipage}[t]{0.47\textwidth}
        \centering
        \adjustbox{valign=c}{\includegraphics[width=0.85\textwidth]{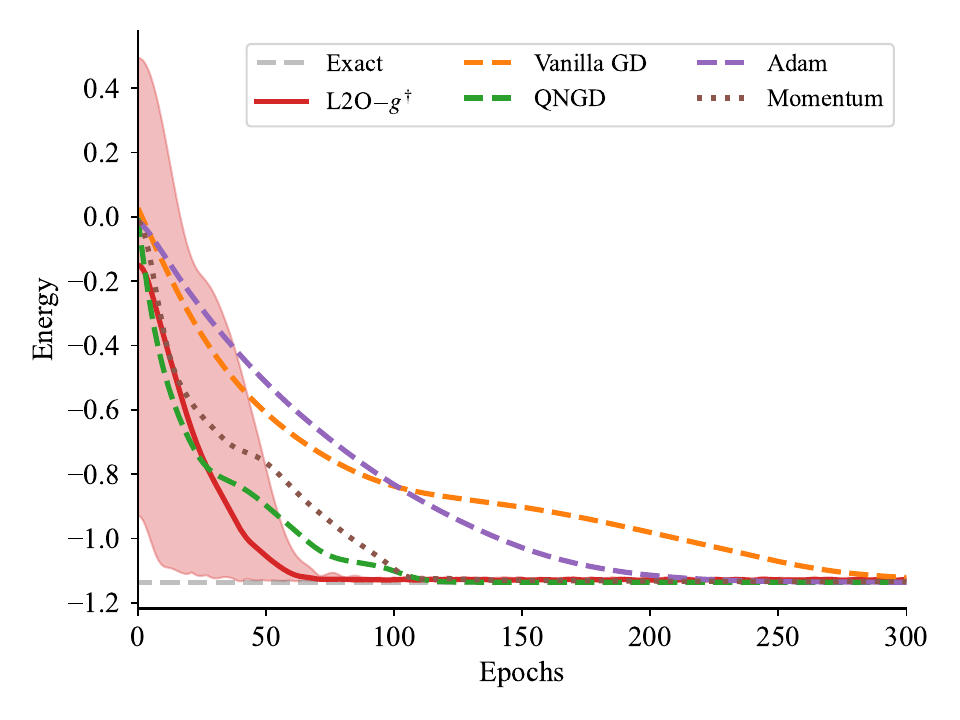}}
        \label{fig:lstm-g_H2}
    \end{minipage}
    \begin{minipage}[t]{0.47\textwidth}
        \centering
        \adjustbox{valign=c}{\includegraphics[width=\textwidth]{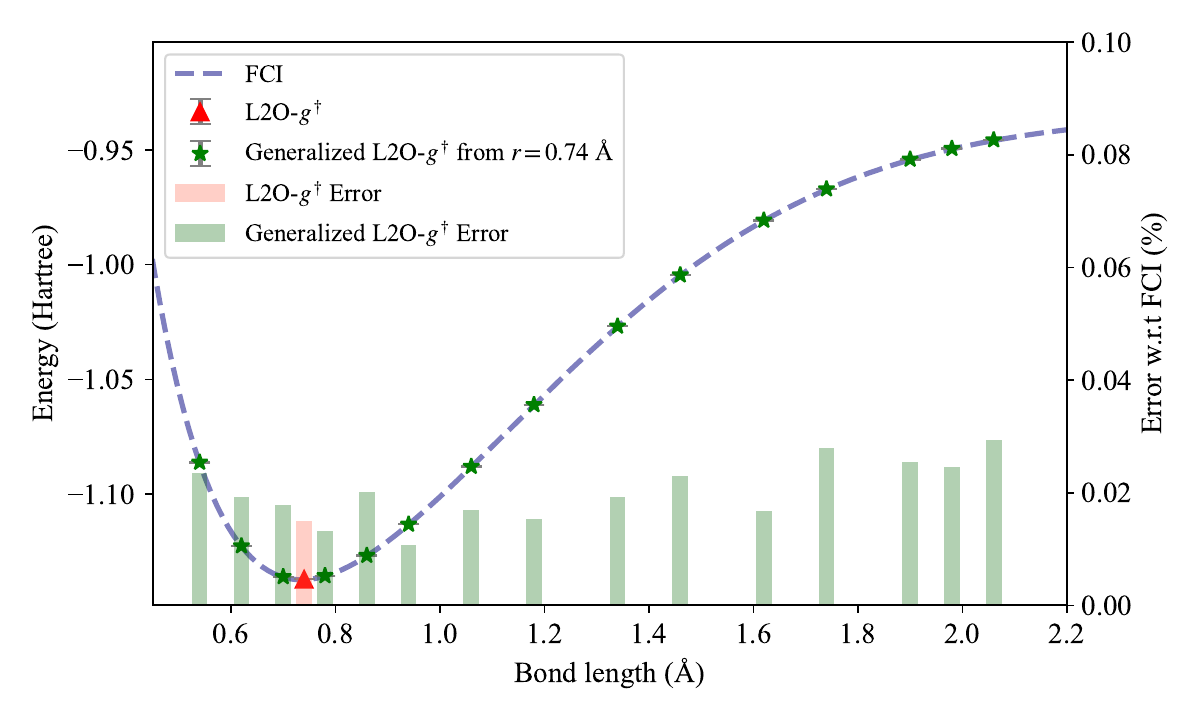}}
        \label{fig:lstm-g_H2_r}
    \end{minipage}
    \caption{\small
    \textbf{(Left) Applying VQE for $\mathrm{H}_2$ using \ours.} 
    We experiment with ten different initial parameters and report the average (line) and standard deviation (ribbon). 
    \ours converges faster than other baseline optimizers for solving the ground state energy of $\mathrm{H}_2$. 
    The grey line is calculated by the FCI model for the exact ground state energy of $\mathrm{H}_2$ at the inter-atomic equilibrium distance \( r=0.7 \, \text{Å} \).
    \textbf{(Right) \ours ability to Generalize on Various Bond Lengths of Optimizing VQE for $\mathrm{H}_2$.} 
    \ours is trained on a single instance from $r=0.74 \text{Å}$ (mark in \textcolor[RGB]{254,42,36}{red}) and then uses the model to optimize various inter-atomic radii (mark in \textcolor[RGB]{8,116,20}{green}). 
    We report the average as well as the standard deviation after 200 iterations of 10 runs and report the relative error to the FCI energy in percentages. 
    }
    \label{fig:vqe_hea}
\end{figure*}

\begin{figure*}[htb!]
    \centering
    \vspace{-1em}
    \includegraphics[scale=0.45]{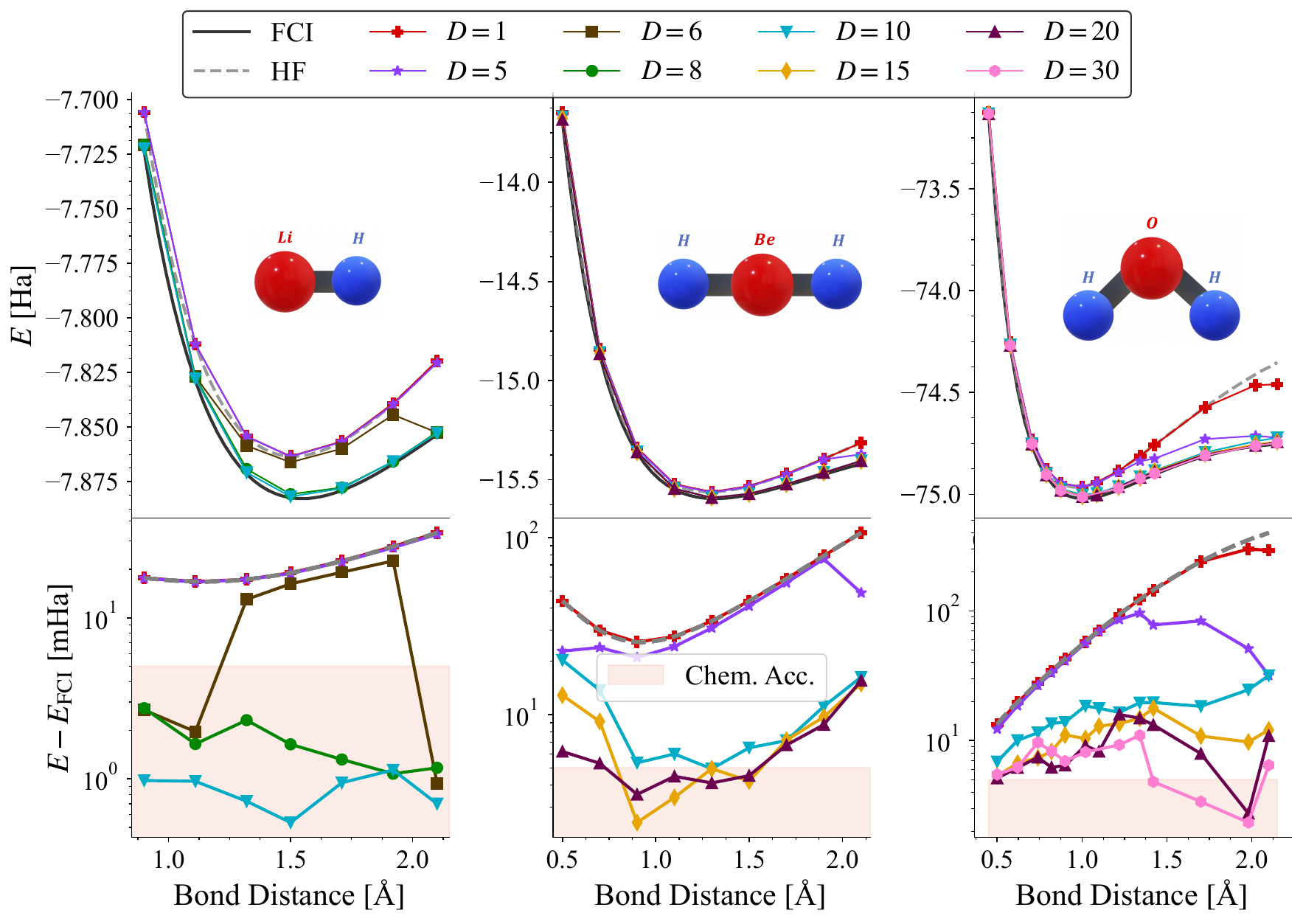}
    \vspace{-1em}
    \caption{\small
    \textbf{Applying VQE with $U^{(1)}_{\text{ent}}$ Ansatz for $\mathrm{LiH}$, $\mathrm{BeH}_2$, and $\mathrm{H}_2\mathrm{O}$ using \ours Trained on Random PQC.}
    For every bond length, we report the average over three random seeds and plot the dissociation profile of various molecules, including $\mathrm{LiH}$, $\mathrm{BeH}_2$, and $\mathrm{H}_2\mathrm{O}$, at different numbers of repeated entangle blocks. 
    As expected, in the lower panel, as we increase the number of entangle blocks, the difference between VQE results and FCI decreases and is even smaller than chemical accuracy in many cases. 
    The chemical accuracy is set to $0.5\times 10^{-2} \text{Ha}$ following \citet{Barkoutsos_2018}. 
    \ours is robust against increasing layers and is even capable of optimizing up to $D=30$.
    }
    \label{fig:vqe_rale}
\end{figure*}

\begin{figure*}[htb!]
    \centering
    \vspace{-1em}
    \includegraphics[scale=0.45]{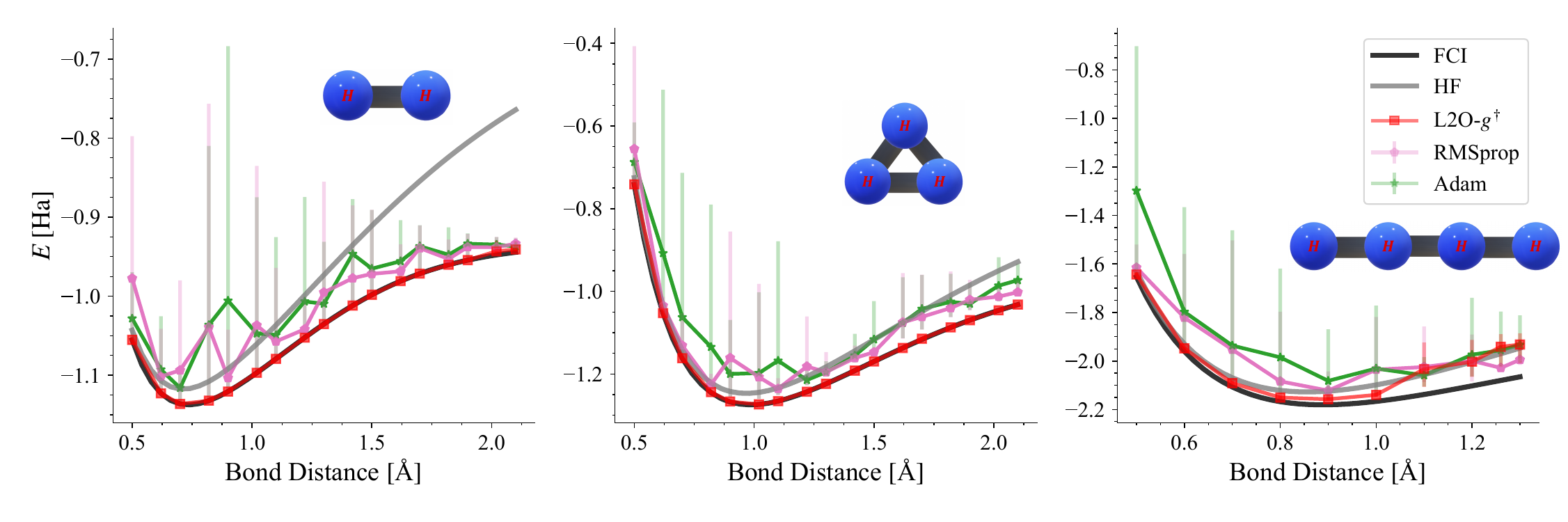}
    \vspace{-1em}
    \caption{\small
    \textbf{Applying VQE with UCCSD Ansatz for $\mathrm{H}_2$, $\mathrm{H}_3^+$, $\mathrm{H}_4$ using \ours Trained on Random PQC.}
    We report the 90\% confidence interval for every bond length along with FCI energy and HF energy.
    The confidence intervals for $\mathrm{H}_2$ and $\mathrm{H}_3^+$ of \ours are too small to be visible.
    }
    
    \label{fig:vqe_uccsd}
\end{figure*}

\paragraph{Results.} 
We illustrate our results on HEA, \( U^{(1)}_{\text{ent}} \) ansatz, and UCCSD ansatz in \Cref{fig:vqe_hea}, \Cref{fig:vqe_rale}, and \Cref{fig:vqe_uccsd}, respectively.
For HEA, shown in \Cref{fig:vqe_hea}, we demonstrate our method's ability by training on a single instance of the \( \mathrm{H}_2 \) molecule to generalize to solve the ground state energy across various radii. 
It surpasses other optimization algorithms in terms of convergence speed.
For the \( U^{(1)}_{\text{ent}} \) ansatz, we show its performance by training on a random generic PQC in \Cref{sec:rand_pqc}. 
It achieves near chemical accuracy on most inter-atomic radii for more complex molecules like \( \mathrm{LiH} \), \( \mathrm{BeH}_2 \), and \( \mathrm{H}_2\mathrm{O} \) with up to 30 entangled blocks.
For the UCCSD ansatz, using the same trained model on the random PQC in \Cref{sec:rand_pqc}, our method achieves higher accuracy compared to RMSprop and Adam optimizers. 
It also shows more robustness against initial parameters, as evidenced by the much smaller 90\% confidence intervals for $\mathrm{H}_2$, $\mathrm{H}_3^+$, and $\mathrm{H}_4$ molecules.

\subsection{QAOA for MaxCut}
\label{sec:maxcut}

\paragraph{Backgrounds.} 
Given an undirected graph \( \mathcal{G} = (\mathcal{V}, \mathcal{E}) \), where \( V \) is the set of vertices, \( E \) is the set of edges, and \( w_{ij} \) is the weight corresponding to the edge \( (i, j) \in E \).
The objective of the MaxCut is to partition the graph vertices \( x_i \), for \( i = 1, \ldots, |V| \) into two sets where the cost function is defined as 
\begin{equation}
    \mathcal{C}(x) = \sum_{i=1}^{|V|} \sum_{j=1}^{|V|} w_{ij} x_i (1 - x_j).
    \label{eq:maxcut_cost}
\end{equation}
MaxCut is $\mathsf{NP}$-hard, meaning no poly-time algorithm solves for all instances assuming $\mathsf{P} \neq \mathsf{NP}$ ($\mathsf{NP}$-Completeness).
Therefore, we rely on approximate solutions. 
We can define the approximation ratio $\alpha$ for an objective function $\mathcal{C}$ as
\begin{equation}
    \alpha = \dfrac{\mathcal{C}(\bm{x}^*)}{\mathcal{C}_{\text{max}}}
    \label{eq:def_alpha}
\end{equation}
where $\mathcal{C}(\bm{x}^*)$ is the value of the objective function for the approximate solution $\bm{x}^*$, and $\mathcal{C}_{\text{max}}$ is the value of the objective function for the optimal solution.
The best-performing classical approximation algorithm for the MaxCut problem is the Goemans-Williamson algorithm, which achieves an approximation ratio of \(\alpha \approx 0.878\).
QAOA was first proposed by \citet{farhi2014quantum}. 
It consists of a parameterized quantum circuit with layer $p_{\text{layer}}$, where each layer includes cost layers $\hat{U}_C(\bm{\gamma}_p)$ and mixer layers $\hat{U}_M(\bm{\beta}_p)$. 
QAOA maps the cost function $\mathcal{C}$ into the Hamiltonian of the circuit. 
It uses variational parameters $(\bm{\gamma}, \bm{\beta})$, which are the cost layer and mixer layer parameters, respectively. 
In total, there are $2p_{\text{layer}}$ parameters in a QAOA problem. 
Regarding the potential of QAOA for the MaxCut problem, \citet{farhi2014quantum} show that for the 3-regular graph MaxCut problem, QAOA achieves a worst-case $\alpha \ge 0.6924$ for $p_{\text{layer}}=1$.

\begin{figure*}[htb!]
    \centering
    \includegraphics[width=0.85\textwidth]{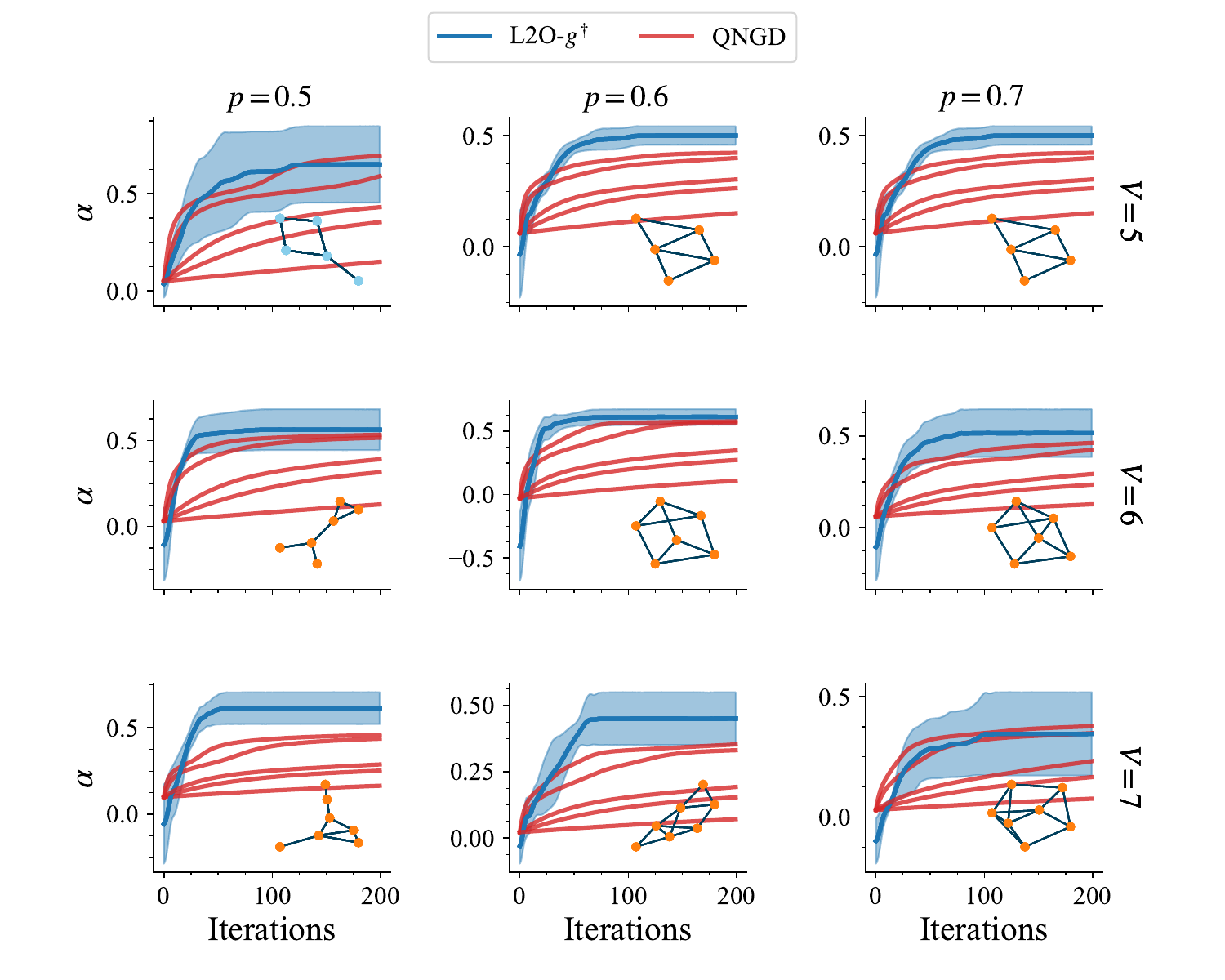}
    \vspace{-1em}
    \caption{\small
    \textbf{Approximation Ratio for the Maxcut Problem with $p_{\text{layer}}=3$ for Randomly Generating the Erdős-Rényi Graph.} 
    We train \ours on the simplest graph $(V,p)=(5,0.5)$ mark in \textcolor[RGB]{124,198,230}{blue}.
    We run the QNGD with five different learning rates, ranging logarithmically from $10^{-2}$ to $10^{-4}$, and averaged the results over five random initial parameters.
   \ours match or surpass QNGD with the best hyperparameter configuration on every graph. We report the results of \ours over the same five random initial parameters and plot the standard deviation in the ribbon. 
   In the cases where QNGD matches \ours happens when the standard deviation is larger, which may be attributed to increased sensitivity to the initial parameters.
   }
    \label{fig:maxcut_all}
\end{figure*}

\paragraph{Setup.}
We construct our map with randomly generated Erdős-Rényi graph $\mathcal{G}(V,p)$ (see \Cref{fig:maxcut_renyi}).
Each pair of $V$ vertices is connected by an edge given with probability \( p \).
The ER graph for different $V$ and $p$ is illustrated in \Cref{fig:maxcut_all}.
We select the ER graph since we can control the difficulty and characteristics of the problem with parameters $V$ and $p$.
We implement the baseline quantum natural gradient descent (QNGD)~\cite{Stokes_2020} with a log-uniform learning rate between $10^{-2}$ to $10^{-4}$ and each learning rate is averaged over five random initial parameters.
For \ours, we only average over the same five random initial parameters since it does not have any hyperparameters and report the mean and standard deviation. 
We train \ours on the simplest graph $(V,p)=(5,0.5)$ mark in blue in \Cref{fig:maxcut_all}.
We set the repeated layer $p_{\text{layer}}=3$ for all graphs.

\paragraph{Results.} 
\Cref{fig:maxcut_all} shows the results for the MaxCut problem on the ER graph with $p \in [0.5, 0.6, 0.7]$ and $V \in [5, 6, 7]$. 
Incidentally, the ER graph coincides for $V=5$ at $p=0.6$ and $p=0.7$.
By training \ours on a simple ER graph, we show that it can match or surpass learning rate-tuned quantum natural gradient descent on various levels of graph complexity.
In the cases where \ours achieves similar results with QNGD, we identify the possible reason for being sensitive to initial parameters.

\subsection{QAOA for Sherrington-Kirkpatrick Model}
\label{sec:sk_model}

\paragraph{Backgrounds.}
Extending the QAOA method from the MaxCut problem in \Cref{sec:maxcut}, in this section we apply QAOA to the Sherrington-Kirkpatrick model.
The SK model~\cite{sherrington1975solvable} is a mean-field model for an all-to-all coupling spin glass.
Consider $n$ Ising spins given by
\begin{equation}
    \sigma = (z_1,\dots,z_n) \in \{ -1,+1 \}^n
\end{equation}
The cost function is defined as 
\begin{equation}
   \mathcal{C}=\dfrac{1}{\sqrt{n}}\sum_{i<j}J_{ij}z_iz_j 
   \label{eq:sk_cost}
\end{equation}
where $J_{ij} = \pm 1$ in our experiment.
Regarding the potential of QAOA for the SK model, \citet{Farhi_2022} demonstrates that the QAOA achieves a higher approximation ratio at \(p=11\) than the classical Semi-Definite Programming (SDP) method.
QAOA showcases the potential for solving the SK model better than the classical SDP method.
Generally, the more layers of the QAOA the more expressibility the circuit becomes, therefore it can achieve lower objective values (cost function).

\paragraph{Setup.}
We test \ours on an $n=8$ SK model, trained on the same ER graph as \Cref{sec:maxcut} with the configuration of $(V,p)=(5,0.5)$.
We select a wide variety of hand-designed optimizers and evaluate them on default hyperparameters (\Cref{app:hyperparameter}), including RMSprop, Adam, GD, QNGD, Momentum, and Adagrad. We experiment with QAOA layers from $p_{\text{layer}}=1$ to $p_{\text{layer}}=5$ and report the average and standard deviation over five runs of the final objective value after 10 and 200 iterations.

\begin{figure*}[htb!]
    \centering
    \includegraphics[width=0.85\textwidth]{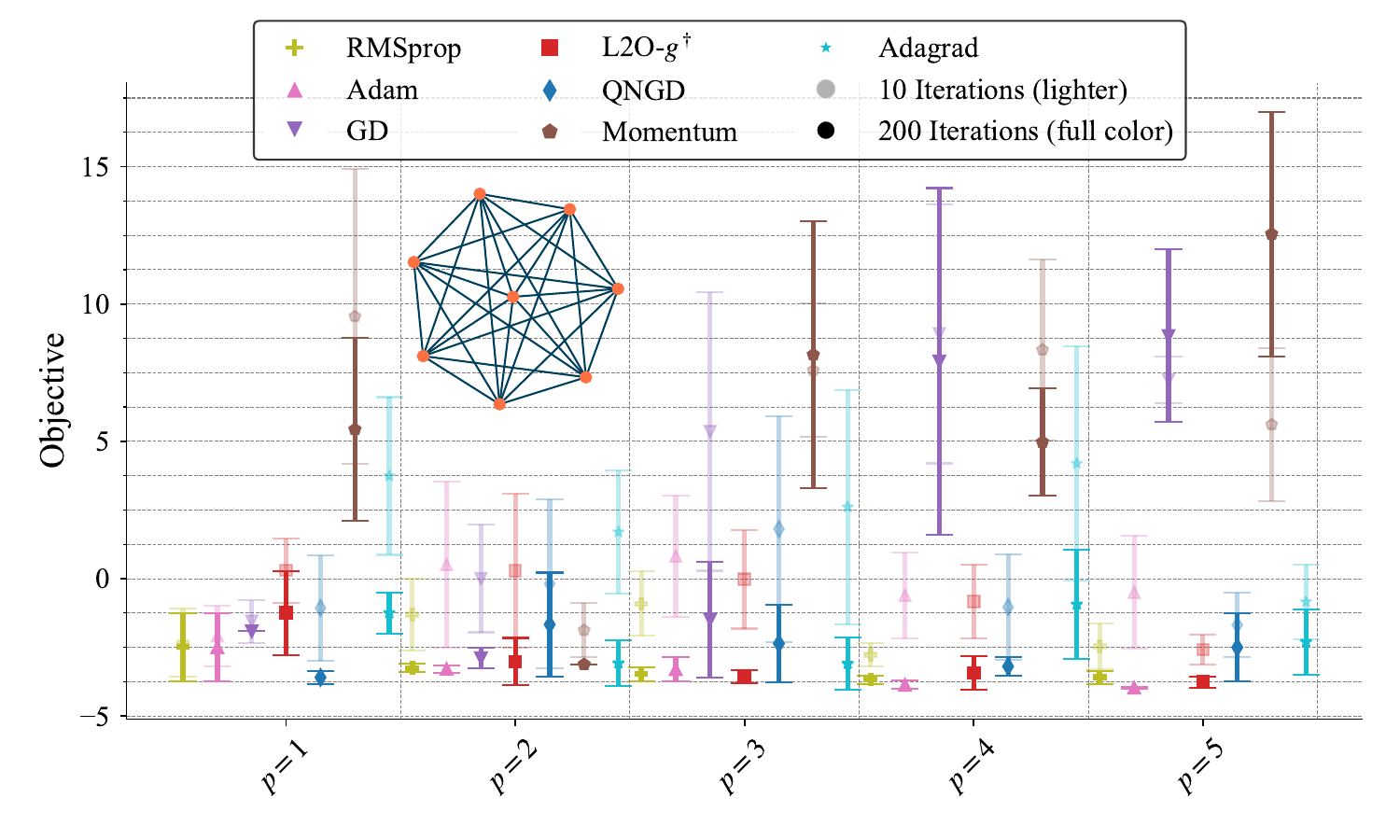}
    \vspace{-1em}
    \caption{\small
    \textbf{Applying QAOA for Sherrington-Kirkpatrick Model with $n=8$ and $p_{\text{layer}} \in [1,5]$.} 
    We plot the final objective values after 10 iterations (lighter color) and after 200 iterations (full color) of various optimizers, including RMSprop, Adam, GD, QNGD, Momentum, and Adagrad, with default hyperparameters. 
    All experiments are repeated five times for every configuration, and we report the mean and standard deviation. 
    \ours consistently converge to lower objective values with relatively small standard deviations, especially converging faster as evidenced by the objective values after ten iterations. 
    }
    \label{fig:sk_model}
\end{figure*}

\paragraph{Results.} 
In \Cref{fig:sk_model}, \ours showcase generalizability and performance by applying QAOA to the SK model after training on a general MaxCut problem of a random ER graph. 
\ours match or surpass hand-designed optimizers without tuning any hyperparameters and converge faster, as evidenced by the final objective value after 10 iterations (lighter color in \Cref{fig:sk_model}).

\subsection{Data Re-upload Classifier}
\label{sec:reupload}

\paragraph{Backgrounds.}
A quantum neural network consists of three major parts: data encoding, state evolution, and measurement. 
First, data encoding takes the input data $\bm{x} \in \mathbb{R}^{x_{dim}}$ and maps it into an $N_q$ qubits Hilbert space with the feature map $|\psi_x\rangle = U_{enc} |0\rangle^{\otimes N_q}$. 
Second, the state evolves with variational unitary $|\psi_{\theta}(x)\rangle = U_{\theta} |\psi_x\rangle$. 
Finally, perform measurement on $N_q$ qubits.
Combining these steps into a single equation, we get:
\begin{align}
|\psi_{\theta}(x)\rangle = U_{\theta} U_{enc} |0\rangle^{\otimes N_q}.
\end{align}

Quantum neural network is the study of data and information with a quantum circuit (see \cite{bowles2024better, Schuld_2022, Biamonte_2017} for more detail).
\citet{abbas2021power} show that quantum neural networks have favorable training landscapes and achieve higher effective dimensions compared to classical neural nets.
The data re-upload classifier is one of them.
Data re-upload classifier was proposed by \citet{P_rez_Salinas_2020}.
They theoretically and experimentally prove that a single qubit is sufficient to construct a universal quantum classifier.
Here we summarize the data-reupload classifier following \citet{bowles2024better}.
Classical neural networks process original data several times for each neuron in the first hidden layers.
On the other hand, the re-upload classifier uses successive, trainable angle embeddings of data. 
For a vector $\bm{x}$ of three features, two trainable three-dimensional real vectors $\bm{w}$ and $\bm{\theta}$, and encodes them as
\begin{equation}
    U(\bm{x} \circ \bm{\omega} + \bm{\theta})
\end{equation}
where
\begin{equation}
    U(\phi) = e^{iZ\phi_1/2}e^{iY\phi_2/2}e^{iZ\phi_3/2}.
\end{equation}
To encode a data $\bm{x}\in \mathbb{R}^d$, we split $x$ into $\left[ d/3 \right]$ vector.
The fidelity cost function is based on the output qubits to one of the states $\vert 0 \rangle$, $\vert 1 \rangle$. $F_{0_j}(\mathbf{x}), F_{1_j}(\mathbf{x})$ is the fidelity of the $j$-th qubit.
The cost function is defined as 
\begin{equation}
    \mathcal{C}(\theta, \omega, \alpha, \bm{x}_i) = \sum_{j=1}^{n_{\text{max}}} (\alpha_{0_j}F_{0_j} - (1 - y_i))^2 + (\alpha_{1_j}F_{1_j} - y_i)^2
\end{equation}
where $\alpha_{0_j}$, $\alpha_{1_j}$ are trainable parameters.
When predicting, we use average fidelity either $\vert 0 \rangle$ or $\vert 1 \rangle$ over the same qubits:
\begin{equation}
    y_{\text{pred}} = \arg \max(\langle F_{0_j} \rangle, \langle F_{1_j} \rangle).
\end{equation}

\paragraph{Setup.}
We test \ours on data re-upload classifier, trained with random PQC from \Cref{sec:rand_pqc}. 
Experiments were conducted using a single qubit classifier across multiple layers $l \in \{ 1, 3, 5, 8, 10 \}$. 
Baseline results are quoted from \citet{P_rez_Salinas_2020} when available, and reproduced otherwise.
Following the experimental setup in \citet{P_rez_Salinas_2020}.
We use a data re-upload classifier to determine whether points lie inside or outside a circle. 
It classifies points within a circular boundary defined by $\bm{x} = (x_1, x_2)$, where $x_i \in [-1, 1]$ and $x_1^2 + x_2^2 < r^2$.
The radius $r$ is set to $\sqrt{2}$. 
We generated a training dataset with 200 datapoints and a testing dataset with 4000 datapoints. 
We then average the results three times to evaluate the model's performance. 
We further conducted 20 random learning rate searches between $10^{-2}$ and $10^{-4}$ for Adam and Momentum and compare them with a single trial \ours.

\begin{figure*}[htb!]
    \centering
    \begin{minipage}[t]{0.50\textwidth}
        \centering
        \adjustbox{valign=c}{\includegraphics[width=0.85\textwidth]{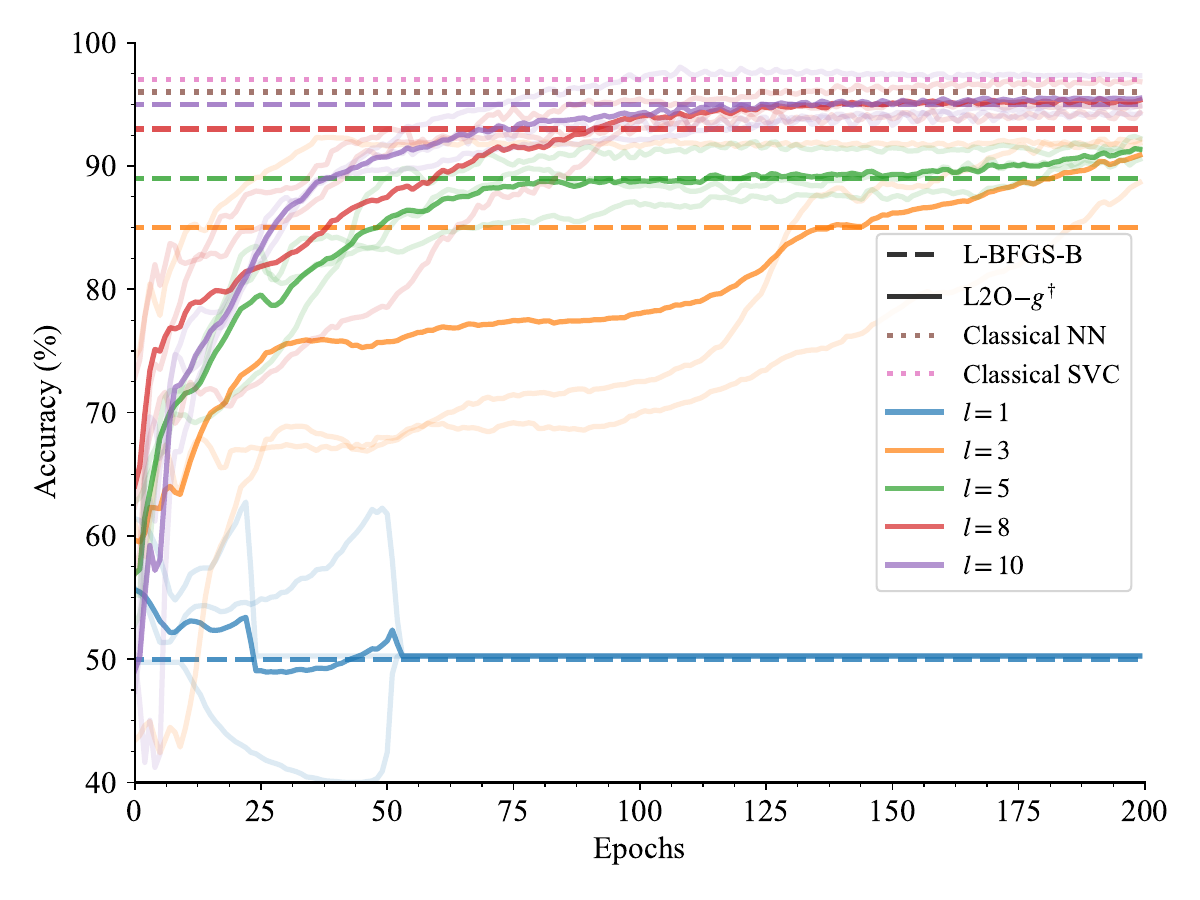}}
    \end{minipage}
    \hspace{-0.05\textwidth}
    \begin{minipage}[t]{0.50\textwidth}
        \centering
        \adjustbox{valign=c}{\includegraphics[width=\textwidth]{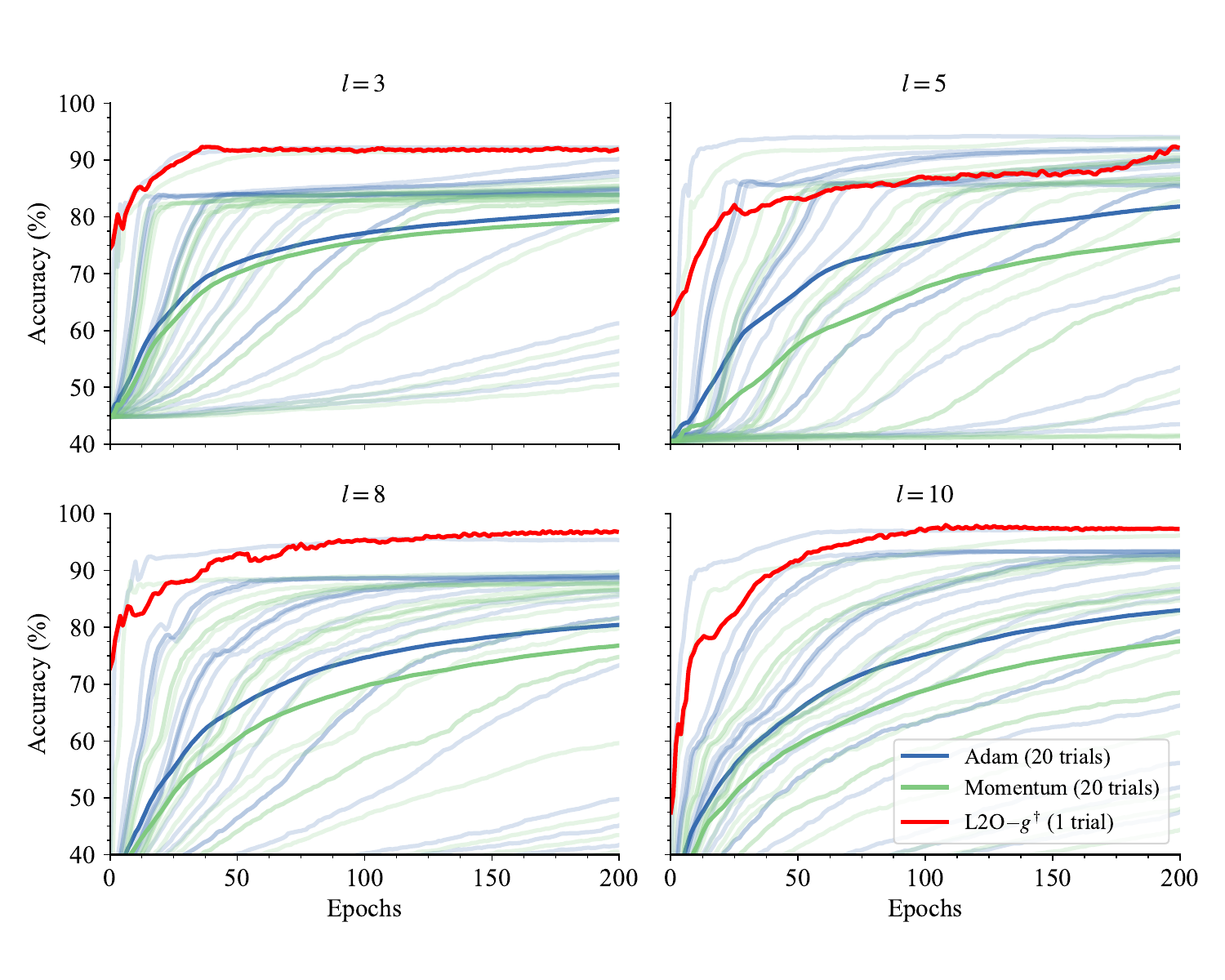}}
    \end{minipage}
    \caption{\small
    \textbf{(Left) Accuracy of \ours on Optimizing Re-upload Circuit on Circular Boundary Classification Problem with $l \in \{ 1, 3, 5, 8, 10 \}$.}
    The results are averaged over three runs and shown in solid-colored lines. 
    The lighter color corresponds to each of the three runs for every configuration. 
    Horizontal dashed lines are quoted from \citet{P_rez_Salinas_2020} using the L-BFGS-B optimizer, a second-order optimizer. 
    The horizontal dotted lines are also quoted from \citet{P_rez_Salinas_2020} for classical neural network results. 
    \ours surpasses the L-BFGS-B optimizer by reaching higher accuracy after a few iterations.
    \textbf{(Right) Compare the Performance of a Single Trial \ours with 20 Random Learning Rate Search Adam and Momentum Optimizer.}
    We average over three runs for every configuration with the same training sets and initial parameters.
    \ours match or surpass 20 random learning rate tuned Adam and Momentum optimizer.
    }
    \label{fig:reupload}
\end{figure*}

\paragraph{Results.}
We summarize the results in \Cref{fig:reupload} and \Cref{tab:reupload}. 
In \Cref{fig:reupload} (Left), we show that replacing the second-order optimizer L-BFGS-B used in \citet{P_rez_Salinas_2020} with \ours boosts the performance of the data re-upload classifier. 
The final accuracy for \ours, L-BFGS-B, gradient descent, Adam, and Momentum are summarized in \Cref{tab:reupload}. 
In \Cref{fig:reupload} (Right), \ours uniformly matches or outperforms 20 random learning rate search Adam optimizer and Momentum optimizer with a single trial \ours optimizer. 
\ours learns a good optimization strategy with minimum effort and showcases convincing results both in terms of optimization wall time and consistency of convergence speed. 
This brings QNNs a step closer to classical neural networks, as evidenced in the case of $l=10$. 
These results highlight that the selection of the optimizer is as crucial as the neural network architecture itself. 
By training on a generic PQC, \ours can tackle drastically different tasks.

\input{tables/reupload}

\subsection{Ablation Studies}
\label{sec:ablation}

We conduct the following ablation studies on \ours:

\paragraph{Compare with LSTM Learned Optimizer.}
In \Cref{tab:ablation}, we compare \ours with L2O-DM~\cite{andrychowicz2016learning}, which has been used with the same architecture in \citet{verdon2019learning, wilson2019optimizing}, utilizing a simple coordinate-wise LSTM. 
Both models are trained on a single random PQC with the configuration of $(N_q, l) = (7, 5)$. 
We randomly select a range from three to six configurations for every problem detailed in \Cref{sec:rand_pqc,sec:vqe,sec:maxcut,sec:sk_model,sec:reupload} and report the average and standard deviation of either loss, approximation ratio, or accuracy over five runs. 
The results show that \ours significantly improves its performance and generalizability from previously learned optimizer despite only being trained on a single generic PQC instance.
\input{tables/ablation}

\begin{figure}[htb!]
    \centering
    \vspace{-1em}
    \includegraphics[scale=0.50]{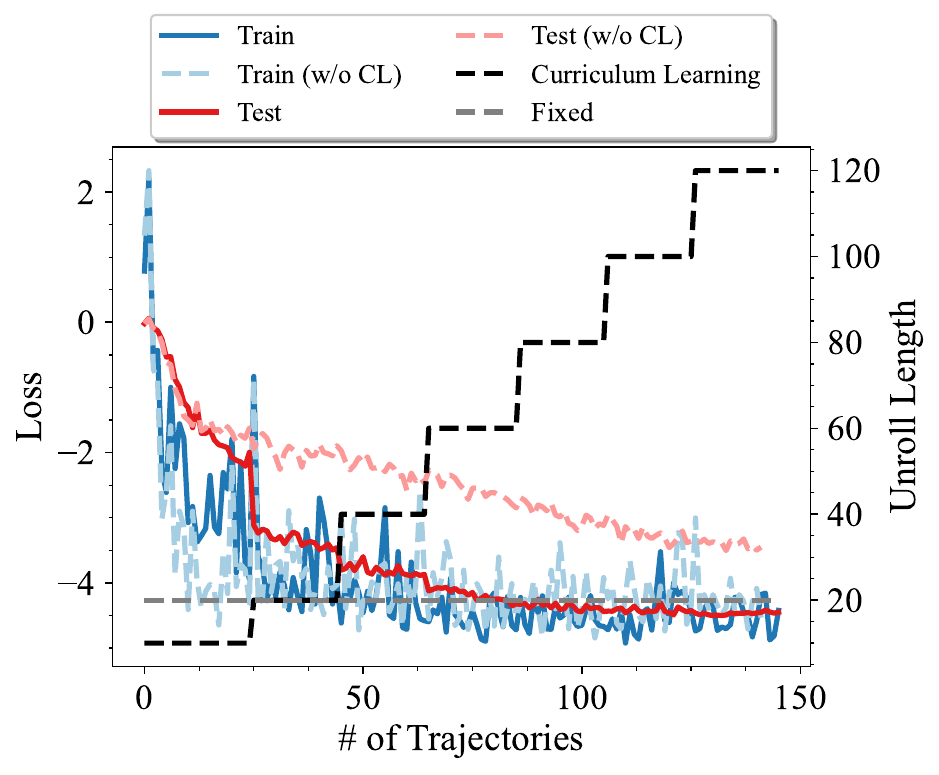}
    \vspace{-1em}
    \caption{\small
    \textbf{\ours Training and Testing Loss with or without Curriculum Learning.}
    The training loss exhibits a similar trend with or without curriculum learning. 
    However, the testing loss for \ours is significantly lower and converges faster than without curriculum learning. 
    The black and grey dashed lines correspond to unroll length after the number of trajectories.
    The y-axis corresponds to the number of trajectories trained by the model.
    }
    \label{fig:train}
\end{figure}

\paragraph{Curriculum Learning.}
In \Cref{fig:train}, we train a separate model without curriculum learning as detailed in \Cref{sec:method}. 
Without curriculum learning, the test loss is significantly higher while the training loss is roughly of the same magnitude as the model using curriculum learning. 
The results highlight that curriculum learning can improve generalizability and provide a more stable training process.
Specifically, \ours only takes roughly 100 trajectories to reach a stable minimum and is able to generalize well on totally different problems (see \Cref{tab:ablation}).

%% file: Tex/random_circuit.tex
\begin{figure}[htb]
    \centering
    \scalebox{0.77}{
    \Qcircuit @C=1em @R=0.7em {
    & \gate{R_y\left(\frac{\pi}{4}\right)} & \gate{R_{P_{1,1}}(\theta_{1,1})} & \ctrl{1} & \qw & \dots & & \gate{R_{P_{1,l}}(\theta_{1,l})} & \ctrl{1} & \qw \\
    & \gate{R_y\left(\frac{\pi}{4}\right)} & \gate{R_{P_{2,1}}(\theta_{2,1})} & \control \qw & \ctrl{1} & \dots & & \gate{R_{P_{2,l}}(\theta_{2,l})} & \control \qw & \ctrl{1} \\
    & \gate{R_y\left(\frac{\pi}{4}\right)} & \gate{R_{P_{3,1}}(\theta_{3,1})} & \qw & \control \qw & \dots & & \gate{R_{P_{3,l}}(\theta_{3,l})} & \qw & \control \qw \\
    & \vdots & & & & \vdots & & & & \vdots \\
    & \gate{R_y\left(\frac{\pi}{4}\right)} & \gate{R_{P_{N_q,1}}(\theta_{N_q,1})} & \qw & \qw & \dots & & \gate{R_{P_{N_q,l}}(\theta_{N_q,l})} & \qw & \qw \\
    }
    }
    \caption{\small
    \textbf{Illustration of Random Parameterized Quantum Circuit.}}
    \label{fig:random_circuit1} 
\end{figure}

%% file: tables/reupload.tex
\begin{table}[ht]
\centering
\begin{sc}
\begin{tabular}{l|lllll} %
\toprule
Layers & 1 & 3 & 5 & 8 & 10 \\ 
\midrule
L-BFGS-B   & 50$^*$ & 85$^*$ & 89$^*$ & 93$^*$ & 95$^*$ \\
Vanilla GD & 44.47  & 49.40 & 48.41 & 43.34 & 47.58 \\
Momentum   & \textbf{56.14}  & 79.35 & 82.47 & 77.57 & 79.25 \\
Adam       & 50.31  & 79.58 & 84.33 & 83.88 & 86.08 \\
\textbf{\ours} & 50.25 & \textbf{90.88} & \textbf{91.37} & \textbf{95.34} & \textbf{95.50} \\
\midrule
C. NN  & & & \textbf{96$^*$} & &  \\
C. SVC & & & \textbf{97$^*$} & & \\
\bottomrule
\end{tabular}
\end{sc}
\caption{\small
\textbf{Accuracy on Various Layers in Data Re-Upload Classifier.} 
We report the average accuracy over three runs.
\ours achieve near optimal score for all layers in data re-upload classifier, even surpassing second-order optimizer L-BFGS-B.
Results quoted from \citet{P_rez_Salinas_2020} are marked with $^*$.
When unavailable, we reproduce optimizer results independently.
}
\label{tab:reupload}
\end{table}

%% file: tables/ablation.tex
\begin{table*}[htb]
\centering
\scalebox{1.0}{
\begin{sc}
\begin{footnotesize}
\begin{tabularx}{0.8\textwidth}{l|l|>{\centering\arraybackslash}X>{\centering\arraybackslash}X}
\toprule
\textbf{Problem} & \textbf{Config.} & \textbf{L2O-DM~\cite{andrychowicz2016learning}} & \textbf{\ours (ours)} \\
\midrule
\multirow{3}{*}{\textbf{Rand. PQC}($\downarrow$)}
& \cellcolor{lightpink}$(N_q,l)=(7,5)$  & \cellcolor{lightpink}\textbf{-1.00 $\pm$ 0.00} & \cellcolor{lightpink}\textbf{-1.00 $\pm$ 0.00} \\
& $(N_q,l)=(7,8)$  & -0.94 $\pm$ 0.06 & \textbf{-1.00 $\pm$ 0.00} \\
& $(N_q,l)=(10,5)$ & \textbf{-1.00 $\pm$ 0.00} & \textbf{-1.00 $\pm$ 0.00} \\
\midrule
\multirow{3}{*}{\textbf{VQE/HEA}($\downarrow$)}
& $\mathrm{H}_2$ at $r=0.5$ & -0.27 $\pm$ 0.31 & \textbf{-0.66 $\pm$ 0.48} \\
& $\mathrm{H}_2$ at $r=0.9$ & -0.52 $\pm$ 0.11 & \textbf{-0.86 $\pm$ 0.21} \\
& $\mathrm{H}_2$ at $r=1.5$ & -0.56 $\pm$ 0.06 & \textbf{-0.93 $\pm$ 0.05} \\
\midrule
\multirow{3}{*}{\textbf{VQE/ENT}($\downarrow$)}
& $\mathrm{LiH}$ at $r=0.9$ & -7.61 $\pm$ 0.03 & \textbf{-7.71 $\pm$ 0.00} \\
& $\mathrm{Be}\mathrm{H}_2$ at $r=0.9$ & -15.23 $\pm$ 0.03 & \textbf{-15.34 $\pm$ 0.00} \\
& $\mathrm{H}_2\mathrm{O}$ at $r=0.9$ & -74.85 $\pm$ 0.01 & \textbf{-74.95 $\pm$ 0.00} \\
\midrule
\multirow{3}{*}{\textbf{VQE/UCCSD}($\downarrow$)}
& $\mathrm{H}_2$ at $r=0.5$ & -0.87 $\pm$ 0.20 & \textbf{-1.06 $\pm$ 0.00} \\
& $\mathrm{H}_2$ at $r=0.9$ & -1.07 $\pm$ 0.06 & \textbf{-1.12 $\pm$ 0.00} \\
& $\mathrm{H}_3^+$ at $r=0.5$ & -0.53 $\pm$ 0.19 & \textbf{-0.74 $\pm$ 0.00} \\
& $\mathrm{H}_3^+$ at $r=0.9$ & -0.79 $\pm$ 0.31 & \textbf{-1.27 $\pm$ 0.00} \\
& $\mathrm{H}_4$ at $r=0.5$ & -0.15 $\pm$ 0.60 & \textbf{-1.47 $\pm$ 0.35} \\
& $\mathrm{H}_4$ at $r=0.9$ & -1.24 $\pm$ 0.21 & \textbf{-2.07 $\pm$ 0.17} \\
\midrule
\multirow{3}{*}{\textbf{MaxCut}($\uparrow$)}
& $V=6, p=0.5$ & 34.78 $\pm$ 12.32 & \textbf{56.36 $\pm$ 11.90} \\
& $V=6, p=0.6$ & 49.06 $\pm$ 9.66 & \textbf{61.56 $\pm$ 6.36} \\
& $V=6, p=0.7$ & 28.67 $\pm$ 9.95 & \textbf{51.53 $\pm$ 12.97} \\
\midrule
\multirow{3}{*}{\textbf{SK Model}($\downarrow$)}
& $V=6, p_{\text{layer}}=1$ & \textbf{-0.93 $\pm$ 0.95} & -0.90 $\pm$ 1.10 \\
& $V=6, p_{\text{layer}}=3$ & -1.95 $\pm$ 1.65 & \textbf{-2.96 $\pm$ 0.06} \\
& $V=6, p_{\text{layer}}=5$ & -1.95 $\pm$ 2.12 & \textbf{-2.96 $\pm$ 0.05} \\
\midrule
\multirow{3}{*}{\textbf{Data Re-upload}($\uparrow$)}
& $l=3$ & 66.87 $\pm$ 1.60 & \textbf{90.08 $\pm$ 1.57} \\
& $l=5$ & 69.50 $\pm$ 7.80 & \textbf{91.36 $\pm$ 0.70} \\
& $l=8$ & 81.97 $\pm$ 3.60 & \textbf{95.34 $\pm$ 1.09} \\
\bottomrule
\end{tabularx}
\end{footnotesize}
\end{sc}
}
\caption{\small
\textbf{Ablation Study on L2O-DM~\cite{andrychowicz2016learning} for Different Problems and Configurations.} 
We report the average and standard deviation of final values after 200 iterations over five runs. 
The \textcolor[RGB]{231,84,128}{pink} column highlights the configurations we train on L2O-DM~\cite{andrychowicz2016learning} as well as \ours.
For Random PQC, VQE/HEA, VQE/UCCSD, VQE/ENT, and the SK Model we report the final loss. 
For the MaxCut problem, we report the final approximation ratio. 
For the Data Re-upload problem, we report the final accuracy.
An arrow pointing downward (\(\downarrow\)) indicates that lower values are better, while an arrow pointing upward (\(\uparrow\)) indicates that higher values are better.
\ours show significantly better generalization away from their training task by converging to lower values.
}
\label{tab:ablation}
\end{table*}

%% file: Tex/4_Related.tex
\section{Related Works}

In this section, we organize related work for deep learning and quantum computing~\Cref{sec:deep}, learned optimizer for neural networks~\Cref{sec:l2o_nn} and learned optimizer for PQC~\Cref{sec:l2o_pqc}.

\subsection{Deep Learning and Quantum Computing}
\label{sec:deep}
Rapid developments in deep learning and quantum computing have drastically widened our imagination about future applications and computing. 
Deep learning is a machine learning model that fits the domains of data~\cite{prince2023understanding}. 
It has shown extraordinary capacity across various domains, including natural language~\cite{achiam2023gpt,team2023gemini}, computer vision~\cite{dosovitskiy2021image,srivastava2023omnivec}, time-series forecasting~\cite{zeng2023transformers,wu2024stanhop, reneau2023feature, zhang2024multivariate}, speech recognition~\cite{ravanelli2021speechbrain,zhang2023google}, tabular learning~\cite{gorishniy2024tabr,xu2024bishop}, and reinforcement learning for control systems~\cite{xu2023beyond}.
In particular, the advent of the transformer-based foundation model in language generation~\cite{hu2024outlier, yu2024enhancingjailbreakattacklarge, luo2024decoupledalignmentrobustplugandplay}, genomics domain~\cite{zhou2023dnabert, zhou2024dnabert} and vision transformer~\cite{hu2024statistical} has significantly advanced the field, showing impressive performance.
Many works combine deep learning with quantum computing in hybrid variational quantum algorithms, including learned optimizers~\cite{verdon2019learning,wilson2019optimizing} and quantum-classical neural networks~\cite{Mari_2020, arthur2022hybrid}.
In the next sections, we organize works that apply deep learning models to either classical neural networks~\Cref{sec:l2o_nn} or parameterized quantum circuits~\Cref{sec:l2o_pqc} to learn to optimize.

\subsection{Learning to Optimize Neural Networks}
\label{sec:l2o_nn}

The first major development for learned optimizers was proposed by \citet{andrychowicz2016learning}, who showed that a simple coordinate-wise LSTM model can outperform hand-designed models on similar tasks from its training sets. 
However, it does not generalize well to different neural network architectures far from its training dataset distribution.
Training techniques are especially important for meta-learning optimizers since the process of back-propagation through a cumulative rolling length makes it unstable. 
\citet{metz2019understanding} show that chaotic dynamics may appear when training a learned optimizer. 
\citet{chen2020training} further incorporate additional features such as validation loss and show that by training on large instances, they can generalize well even when training themselves from scratch. 
Several researchers have identified the importance of training techniques for learned optimizers and empirically show improvements in the variety of SOTA optimizers. 
For example, \citet{chen2020training} use curriculum learning and imitation learning to improve training stability and generalizability; 
\citet{verdon2019learning} propose random scaling and incorporate convex functions to improve the generalizability of learned optimizers. 
Instead of using a black-box model, \citet{chen2023symbolic} use symbolic learning to discover new optimization algorithms. 
\citet{metz2022velo} scale up the training sets for the learned optimizer and use extensive training on a wide variety of deep learning problems, showing that the learned optimizer demonstrates superior performance in terms of time saved for hyperparameter optimization.

\subsection{Learning to Optimize Parameterized Quantum Circuits}
\label{sec:l2o_pqc}

\citet{verdon2019learning,wilson2019optimizing} utilized same architecture as \citet{andrychowicz2016learning}.
However, it can only serve as an initializer and is limited to similar optimization tasks in their training sets.
Another work by \citet{kulshrestha2023learning} also uses a similar architecture to \citet{andrychowicz2016learning}, they use the change in the cost function to approximate the gradient as the input feature. 
Their tasks are limited to a fixed PQC for quantum machine learning tasks.
Other lines of work by \citet{lockwood2022optimizing} utilizing Reinforcement Learning (RL) to augment gradient descent algorithm.
Their results show that the performance critically depends on the structure of the ansatz.
Previous studies on learned optimizers for PQC demonstrate their ability to optimize VQAs in limited scenarios outside the training distribution.

%% file: Tex/6_Conclusion.tex
\section{Conclusion}
In this paper, we present \ours, a \textit{quantum-aware} learned optimizer that can generalize to a wide variety of parameterized quantum circuits by leveraging parameter space and distribution space optimization. 
\ours is tailored for optimizing parameterized quantum circuits, requiring minimal training instances, and experimentally shows strong out-of-distribution generalization with the ability to optimize diverse VQAs like VQE, QAOA, and QNN. 
Importantly, we demonstrate that out-of-the-box \ours can match or surpass learning rate-tuned optimizers like Adam and QNGD, as well as previous work on learned optimizers for PQC, in terms of generalization and convergence speed.

However, implementing VQAs on a large scale in the NISQ era poses additional challenges for surpassing classical computers. 
These challenges include trainability, efficiency, error mitigation, and scaling up VQAs. 
Our work takes a small step toward addressing the intrinsic hard problems of optimizing VQAs.

\section{Limitations}
\label{sec:limit}
The current learned optimizer architecture is quite simplistic. 
Incorporating more advanced deep learning models~\cite{cao2019learning, metz2022practical, metz2022velo} and training techniques~\cite{wichrowska2017learned, metz2019understanding} into the quantum-aware learned optimizer are crucial future directions.
While we strive to benchmark a wide variety of cases on \ours to test the out-of-distribution performance from training on a single PQC instance, there may still be instances where the learned optimizer struggles. 
Additionally, an important direction for future work is to scale up the training sets, as highlighted in \citet{metz2022velo}, to thoroughly test the scalability of the learned optimizer.

%% file: Tex/Appendix.tex
\input{Tex/App_exp}

%% file: Tex/App_exp.tex
\section{Erdős-Rényi graph}
\label{sec:ER_graph}
\begin{figure}[H]
    \centering
    \vspace{-1em}
    \includegraphics[scale=0.25]{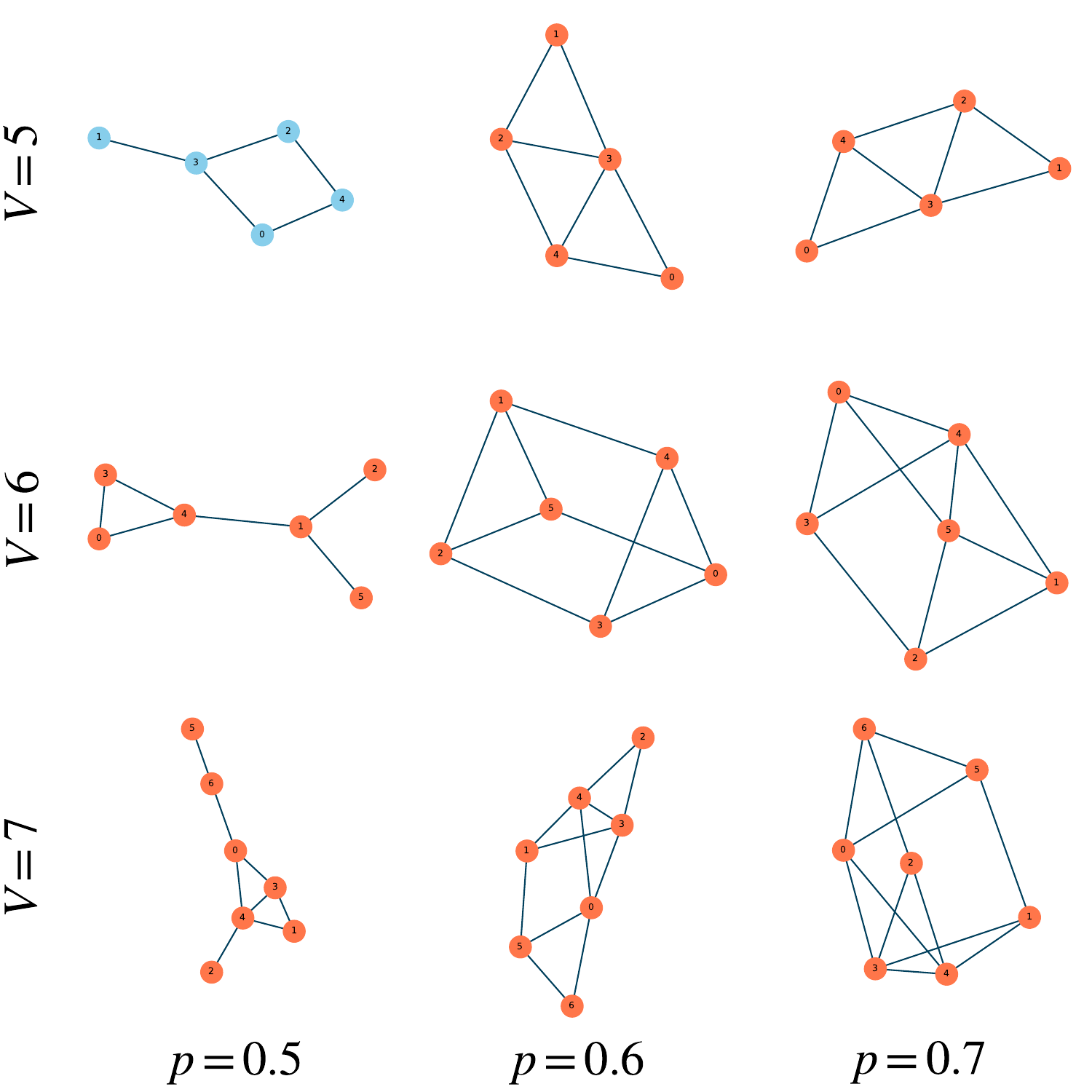}
    \caption{
    \textbf{Examples of Randomly Generated Erdős-Rényi $\mathcal{G}(n,p)$.}
    We randomly generate the Erdős-Rényi graph for the MaxCut problem for vertices $V \in \{ 5, 6, 7\}$ and probability $p \in \{ 0.5, 0.6, 0.7 \}$  where $n$ is the number of vertices and each pair of $n$ vertices connected by an edge with probability \( p \). 
    Horizontal direction represents different $p$ values and vertical direction represents different values of $V$.}
    \label{fig:maxcut_renyi}
\end{figure}

\section{Hyperparameter Space}
\label{app:hyperparameter}
\input{tables/hpo_space.tex}

\section{Version Information}
\label{sec:version}
\input{tables/version}

%% file: tables/hpo_space.tex
\begin{table}[htb!]
    \centering
    \caption{\small
    \textbf{Default Hyperparameters for Baseline Optimizers.}
    }
    \begin{tabular}{lll}
    \toprule
    Optimizer & Default lr & Default Values \\
    \midrule
    Adam & 0.01 & $\beta_1=0.9, \beta_2=0.999, \epsilon=10^{-8}$ \\
    Adagrad & 0.01 & $\epsilon=10^{-8}$ \\
    RMSprop & 0.01 & $\rho=0.9, \epsilon=10^{-8}$ \\
    Momentum & 0.01 & $m=0.9$ \\
    Vanilla GD & 0.01 & N/A \\
    QNGD & 0.01 & $\lambda=0.01$ \\
    \bottomrule
    \end{tabular}
\end{table}

%% file: tables/version.tex
\begin{table}[H]
\centering
\caption{\textbf{Software and System Information.}}
\begin{tabular}{ll}
\toprule
Software & Version \\
\midrule
pennylane & 0.34.0 \\
wandb & 0.16.3 \\
tqdm & 4.66.2 \\
scikit-learn & 1.4.1.post1 \\
matplotlib & 3.8.3 \\
torch & 4.66.2 \\
torchvision & 0.17.2 \\
torchaudio & 0.10.0+rocm4.1 \\
h5py & 3.10.0 \\
fsspec & 2024.2.0 \\
aiohttp & 3.9.3 \\
\midrule
Parameter & Value \\
\midrule
Python version & 3.9.18 \\
Python compiler & GCC 11.2.0 \\
Python build & main, Sep 11 2023 13:41:44 \\
OS & Linux \\
CPUs & 40 \\
CPUs Memory (GB) & 754 \\
GPUs & 4 (Tesla V100S-PCIE-32GB) \\
GPUs Memory (GB) & 128 \\
\midrule
\multicolumn{2}{c}{Fri Apr 19 03:05:51 2024 UTC} \\
\bottomrule
\end{tabular}
\label{tab:version}
\end{table}